\documentclass[conference]{IEEEtran}

\usepackage{cite}
\usepackage{amsmath,amssymb,amsfonts}
\usepackage{algorithmic}
\usepackage{graphicx}
\usepackage{textcomp}
\usepackage{xcolor}
\usepackage[hyphens]{url}
\usepackage{fancyhdr}
\usepackage[bookmarks=true,breaklinks=true,colorlinks,citecolor=blue,linkcolor=blue,urlcolor=blue]{hyperref}

\fancypagestyle{firstpage}{
  \fancyhf{}
  
  \fancyfoot[C]{\thepage}
}

\title{\zynqparrot{}: A Scale-Down Approach to Cycle-Accurate, FPGA-Accelerated Co-Emulation} 
\author{
  Daniel Ruelas-Petrisko (University of Washington) <petrisko@cs.washington.edu> \\
  Farzam Gilani (University of Washington) <farzamgl@uw.edu> \\
  Anoop Mysore Nataraja (University of Washington) <mysanoop@uw.edu> \\
  Zoe Taylor (Seattle Country Day School) <zoe.nguyen.taylor@gmail.com> \\
  Michael Taylor (University of Washington) <prof.taylor@gmail.com>
}

\newcommand{\blackparrot}{BlackParrot}

\newcommand{\zynqparrot}{ZynqParrot}
\newcommand{\projectname}{\zynqparrot{}}
\newcommand{\panicroom}{PanicRoom}
\newcommand{\plshell}{P-Shell}
\newcommand{\nbfifo}{SB-FIFO}
\newcommand{\zynqps}{VPS}
\newcommand{\farm}{ZP-Farm}
\newcommand{\catchup}{Catch-up}
\newcommand{\scaledown}{Scale-Down}

\newcommand{\zynqparroturl}{https://github.com/black-parrot-hdk/zynq-parrot}

\newcommand{\panicroomurl}{https://github.com/black-parrot-sdk/libgloss-dramfs}

\newcounter{todocounter}
\newcommand{\bsgtimes}{$\times$}

\usepackage{siunitx}
\usepackage{tikz} 
\usetikzlibrary{angles, automata, backgrounds, positioning, arrows.meta, quotes, shapes, calc}
\usepackage[caption=false]{subfig} 
\usepackage[T1]{fontenc} 
\usepackage{threeparttable} 
\usepackage{etoolbox} 
\apptocmd{\sloppy}{\hbadness 10000\relax}{}{}
\tikzset{
    every picture/.append style={node distance=4cm, on grid, auto, thick},
    every node/.append style={outer sep=0pt, align=center},
    every edge/.append style={},
    bsg_circle/.style ={circle, draw, fill=white},
    bsg_rectangle/.style={rectangle, draw, fill=white},
    bsg_cloud/.style={cloud, draw, fill=white},
    bsg_triangle/.style={draw, isosceles triangle},
    bsg_trapezoid/.style={draw, trapezium, trapezium angle=45},
    >=latex
}
\usepackage{rotating}               
\usepackage{adjustbox}              
\usepackage{verbatim} 

\usepackage{listings}
\definecolor{codegreen}{rgb}{0,0.7,0}
\definecolor{codegray}{rgb}{0.5,0.5,0.5}
\definecolor{codepurple}{rgb}{0.58,0,0.82}
\definecolor{codered}{rgb}{1,0,0}
\definecolor{backcolour}{rgb}{0.94,0.94,0.94}
\lstdefinestyle{cppstyle}{
  backgroundcolor=\color{backcolour},
  language=[11]C++,
  commentstyle=\color{codegreen},
  keywordstyle=\color{magenta},
  numberstyle=\tiny\color{codegray},
  stringstyle=\color{codepurple},
  basicstyle=\ttfamily\scriptsize,
  breakatwhitespace=false,
  breaklines=true,
  captionpos=b,
  keepspaces=true,
  numbers=left,
  showspaces=false,
  showstringspaces=false,
  showtabs=false,
  tabsize=2,
  frame = tblr,
  framesep = 3pt,
  framerule = 0.4pt
}

\usepackage{multicol} 
\usepackage{bbding} 
\usepackage{csvsimple} 

\usepackage{sepfootnotes}
\begin{document}
\maketitle
\thispagestyle{firstpage}
\pagestyle{plain}
\begin{abstract}
As processors increase in complexity, costs grow even more rapidly, both for functional verification and performance validation. Additionally, performance models become ever more sensitive to slight microarchitecture inaccuracies. Runtime measurements of key workloads are an essential part of the performance debugging process. Most often, silicon characterizations comprise simple performance counters, which are aggregated and separated to tell a story. Based on these inferences, performance engineers employ microarchitectural simulation to inspect deeply into the core. Unfortunately, dramatically longer runtimes make simulation infeasible for long workloads.

Traditionally, architects have bridged this gap by performing early prototyping on FPGA. Yet, the scale of modern designs is impractical to implement on a single emulation board. Large companies use Scale-Up solutions such as commercial emulation platforms, but these are unaffordable to academics, hobbyists and startups. Others have proposed Scale-Out solutions, leveraging cloud FPGA clusters to emulate large System-On-Chips. However, this approach prescribes certain I/O and memory system architectures instead of the native interface timings of any given subsystem.

Instead, we propose a \textit{Scale-Down} approach to modelling and validation. Rather than up-sizing a prototyping platform to fit large and complex system designs, we show that it can be more accurate, faster, and more economical to decompose a system into manageable sub-components that can be prototyped independently. By carefully designing the prototyping interface, it is possible to adhere to strict non-interference of the Device Under Test (DUT). This allows architects to have the best of both worlds: the speed of FPGA acceleration while eliminating the inaccuracies of Scale-Out and the inherent costs of Scale-Up.

In this work, we present \projectname{}: a Scale-Down FPGA-based modelling platform, capable of executing non-interfering, cycle-accurate co-emulations of arbitrary RTL designs. \projectname{} is capable of verifying functionality and performance with arbitrary granularity. We also provide case studies using \projectname{} to analyse the full-stack performance of an open-source RISC-V processor.

\end{abstract}

\maketitle
\thispagestyle{firstpage}
\pagestyle{plain}

\section{Introduction}

As processors increase in complexity, verification costs grow even more rapidly, both for functionality and performance. The end of Dennard Scaling~\cite{4785534} has led to a Cambrian explosion of domain-specific accelerators which present unique, full-stack verification challenges. Besides functional correctness, performance validation is critical to achieving worthwhile gains from accelerator integration. The higher performance the design, the more sensitive it is to subtle disturbances in the microarchitecture. When characterizing full-system performance in silicon, software engineers use simple performance counters which must be decided upon early in the design process, before problematic subsystems have been identified. These counters are aggregated and separated to divine reasons for the performance of a target application. On the other end of the spectrum, architects can leverage the deep verification capabilities of pre-silicon waveform inspection to identify subsystem bottlenecks before tapeout, when fixes are much cheaper. However, cycle-accurate simulations are painfully slow, so architects must settle for sampling applications~\cite{hamerly2005simpoint} to complete in a reasonable timescale.

Traditionally, architects have bridged this gap by performing early prototyping in FPGA. By doing so, RTL similar to tapeout designs can be emulated with cycle accuracy at 10-100x faster than simulation alone. However, modern designs are too large to economically fit on a single emulation board. Large companies can Scale-Up their prototyping systems using commercial emulation platforms~\cite{zebu,palladium,veloce}, but these are unaffordable to academics, hobbyists and startups. Other academics have proposed Scale-Out solutions that leverage cloud FPGA clusters~\cite{aws,alibaba} to emulate large System-On-Chips. However, this approach generally relies on regularity in the design, prescribes certain standardized I/O and memory system architectures and couples platforms to proprietary vendor IP which may interfere with the native interface timings of any given system and provide no insight or ability to adapt for system needs.

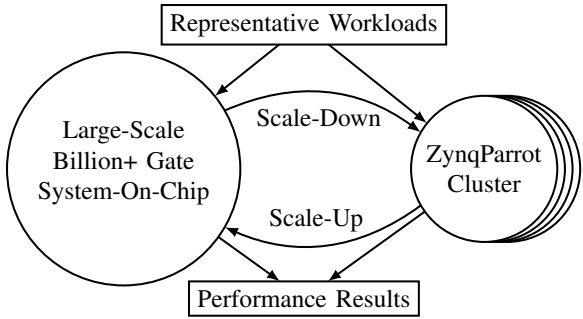
\begin{figure}
    \centering
    \resizebox{8cm}{!}{
    \begin{tikzpicture}
    \node[bsg_circle,                              ] (n0)  {\\Large-Scale\\Billion+ Gate\\System-On-Chip\\};
    \node[bsg_rectangle,          above right=2cm and 2.5cm of n0] (n1)  {Representative Workloads};
    \node[bsg_circle, xshift=9px, below right=2cm and 2.5cm of n1] (n2c) {\projectname{}\\Cluster};
    \node[bsg_circle, xshift=6px, below right=2cm and 2.5cm of n1] (n2b) {\projectname{}\\Cluster};
    \node[bsg_circle, xshift=3px, below right=2cm and 2.5cm of n1] (n2a) {\projectname{}\\Cluster};
    \node[bsg_circle,             below right=2cm and 2.5cm of n1] (n2)  {\projectname{}\\Cluster};
    \node[bsg_rectangle,          below=3.8cm of n1] (n3)  {Performance Results};

\path [->]
    (n0)    edge    (n3)
    (n1)    edge    (n0)
    (n1)    edge    (n2)
    (n2)    edge    (n3)
    (n0)    edge[bend left, swap]
            node[xshift=25px, yshift=-3px]{Scale-Down}
                    (n2)
    (n2)    edge[bend left, swap]
            node[xshift=18px, yshift= 2px]{Scale-Up}
                    (n0);
\end{tikzpicture}
    }
    \caption{
    A Scale-Up/Scale-Down cycle transforms industrial manufacturing processes into laboratory experiments. Experiments accept identical inputs as the full process. Results are extrapolated to predict impacts on the modified process before committing to costly change orders. Iterating over the process lifecycle leads to continuous improvement.
    }
    \label{fig:01-process}
\end{figure}

Inspired by biotechnological process modelling~\cite{mirasol2019scaledown}, we propose a \textit{Scale-Down} approach for architectural prototyping, shown in Figure~\ref{fig:01-process}. Instead of unilaterally scaling up a chip design (process) from an FPGA prototype (lab-scale experiment) to a full tapeout (industrial manufacturing process), it is more economical to iteratively Scale-Up and Scale-Down the design to identify subsystem issues at scale and precisely debug performance in a smaller, more tightly instrumented system. By closely correlating the two environments, Scale-Down results can give deep debugging insights into process deviations as well as accurately predict future deviations for orders of magnitude lower cost than a full production run.

In this work, we show that decomposing and recomposing the system is more accurate, faster, and more economical than state of-the-art alternatives. By carefully designing the platform interfaces, it is possible to provide flexible environments that represent the interactions of real systems while maintaining system timings at the component interfaces. This allows architects to have the best of both words: the speed of FPGA acceleration while eliminating the inaccuracies of Scale-Out and the inherent costs of Scale-Up. Because iteration time is much faster than monolithic prototypes, small design teams can quickly bootstrap new Scale-Down subsystems, run large simulations on abstracted Scale-Up models and return to Scale-Down to rapidly debug and enhance components.

When Scaling-Down, great care must be taken to accurately transform the design and exactly mimic the environment between the full design and the subsystem. In particular, I/O interface timings must adhere to accurate timing models. Previous works~\cite{karandikar2018firesim, chirkov2023smappic} have focused on instrumenting FPGA timing for common interfaces such as DRAM and Ethernet rather than the fully custom models required for subsystem partitioning. Others~\cite{renode} have focused on providing full-system emulation for software development and too high a level of abstraction for microarchitectural debugging. Instead, this work aims to provide more precise and more flexible interface emulation to allow for finer-grained partitioning without sacrificing accuracy.

Previous FPGA emulation platforms~\cite{zebu}~\cite{palladium}~\cite{karandikar2018firesim}~\cite{chirkov2023smappic} are expensive, dependent on vendor IP, or cumbersome and prone to lock-up. In contrast, \projectname{} builds upon the BaseJump STL~\cite{taylor2018basejump} library to provide generic and completely open bridges to commonly available AXI and UART interfaces. When using Zynq-based FPGAs~\cite{zynq}, the only requirement to use \projectname{} is an SSH-capable machine running Vivado. For non-Zynq FPGAs, \projectname{} requires a UART connection to the FPGA as well as a JTAG connection for bitstream programming, although the platform architecture easily supports plugins for additional host functionality. Contrast this to traditional solutions which require expensive PCIe-capable accelerators. These setups are hard to maintain, built on top of proprietary PCIe IP and software layers such as Xilinx XDMA~\cite{xdma}. Failure to interface correctly to PCIe can lockup not only the DUT but also the host server machine, requiring remote restart capabilities. A cluster of such host machines and FPGAs can easily exceed 10s of thousands of dollars and require full-time system administration (see Table~\ref{tab-01-introduction:comparison}). In contrast, \projectname{} provides verification teams the ability to begin with the minimal possible Total Cost of Ownership (\textit{TCO}) and scale costs alongside the design progress.

\begin{table}
    \centering
    \caption{On a per-FPGA basis Scale-Down systems require a much smaller investment, allowing teams to incrementally build up their verification infrastructure. 
    }
    \begin{threeparttable}
    \small
\begin{tabular}{llll}
\hline
\textbf{Strategy}                                                                                        & \textbf{\$/year/FPGA\tnote{0}}    & \textbf{Logic Unit}        & \textbf{Required I/O}                      \\ \hline
Scale-Up\tnote{1}                                                                       & $\sim$\$5000         & Full Design       & Native PCIe                                        \\ 
Scale-Out\tnote{2}                                                                      & $\sim$\$2000         & Tile              & PCIe Tunnel                                        \\
\textbf{\begin{tabular}[c]{@{}l@{}}Scale-Down\tnote{2}\\\end{tabular}} & \textbf{$\sim$\$100} & \textbf{IP Block} & \textbf{SSH/Serial} \\
\end{tabular}
    \begin{tablenotes}
        \footnotesize
        \item[0] 2000~hours is equivalent to a year of 8-hour regressions.
        \item[1] \$1.4167 per AWS f1.16xlarge hour~\cite{awspricing}.
        \item[2] \$0.6744 per AWS f1.4xlarge hour~\cite{awspricing}.
        \item[3] \$300 per Avnet Ultra96v2 board~\cite{ultra96pricing}, with a replacement rate of once per three years. Cluster MTBF is 100+ years~\cite{ug116}.
    \end{tablenotes}
    \end{threeparttable}
    \label{tab-01-introduction:comparison}
\end{table}

\projectname{} adheres to strict non-interference of internal design timings through strategic clock-gating during unpredictable host back-pressure. This allows subsystems to execute with the illusion that they are running in situ within the full system. Each Device-Under-Test (DUT) execution cycle can be verified against a emulated model, taking into account functional correctness along with verification of internal and external performance. Additionally, \projectname{} is able to synthesize complex performance counters without interfering with mature or frozen RTL. Such a deep dissection of the subsystem can allow an architect to design experiments to identify subsystem bottlenecks and quickly iterate without requiring microarchitectural changes.

We provide case studies of how to analyse a complex RISC-V processor, \textit{\blackparrot{}}~\cite{petrisko2020blackparrot}, using \projectname{}. We identify useful software and hardware enhancements to quickly verify functionality and performance and how they easily fit into the \projectname{} framework. These non-invasive, instrumented measurements and host software abstraction layers were written for use in \blackparrot{}; however, they are generally applicable to a wide range of RISC-V projects. Additionally, we demonstrate real-world uses of \projectname{}: first to host the term projects of an undergraduate/graduate architecture class of 20 students, second to bring up first-batch silicon during a commercial tapeout, and third to analyse a bottleneck and access the improvement of a new microarchitectural widget in \blackparrot{}.

\sepfootnotecontent{1}{The hardware and software code for \projectname{} is open-source under a permissive BSD-3 License. (\zynqparroturl{})}
\sepfootnotecontent{2}{The software code for \panicroom{} is open-source and as of submission time being actively upstreamed to the Newlib project (\panicroomurl{})}

Our major contributions are:
\begin{enumerate}
    \item We present \projectname{}\sepfootnote{1}, the first Scale-Down open-source prototyping platform for deterministic, cycle-accurate local FPGA co-emulation.
    \item We provide a co-simulation capable prototyping library that eschews vendor dependencies and is cycle-identical to \projectname{} co-emulation.
    \item We explore the diversity of \projectname usage via a series of case studies, from the classroom to academic research to commercial bring-up.
    \item We introduce \panicroom{}\sepfootnote{2}: an ultra-portable library that leverages DRAM to run POSIX benchmarks on bare-metal systems.
\end{enumerate}

\section{\projectname{} Architecture}

Complex IP designs often require specific Vivado versions to ensure reproducible builds. However, each Vivado version only supports a small subset of operating systems, often directly conflicting with strict ASIC EDA vendor compatibility guides. Although Vivado-based build systems can be broken up into bitstream generation and programming machines, this requires supplemental system administration.

Additionally, many FPGAs are only available in PCIe-attached variants, resulting in complex and fragile bridge solutions. These connections require proprietary vendor solutions such as Xilinx XDMA~\cite{xdma}, non-portable RTL and software layers and are prone to full-system lockup due to subtle bugs in configuration or execution. Lockup can cause long-running experiments to be lost, force costly manual debugging and complicate systems with fallback-recovery mechanisms. 

Considering these limitations, emulation management software is often tailored to each specific host-FPGA pair and disparate from simulation-based testbench infrastructure, which must be maintained independently. As a result, to normalize environments across designs for powerful prototyping boards, groups must maintain costly heterogeneous server infrastructures with multiple x86 environments customized to each design.

Traditional FPGA prototyping systems require powerful discrete host servers to:
\begin{enumerate}
    \item Compile a netlist and generate a bitstream, often requiring hours for large designs.
    \item Program the bitstream, typically over a USB/JTAG connection.
    \item Manage emulation execution, running software to load test vectors, monitor performance or verify results.
\end{enumerate}

In this section, we describe the \projectname{} architecture and how it addresses these challenges cost-effectively and with lower maintenance than previous solutions.

\subsection{Hardware Architecture}

\begin{figure*}
    \centering
    \subfloat[
        \projectname{} subcomponents interface with the \plshell{} through a parameterizable set of \nbfifo{}s and CSRs. The \plshell{} logic is run asynchronously to the DUT, allowing for decoupled co-emulation. Clock gating logic on the \zynqps{} side ensures accurate co-emulation by maintaining internal timings of the DUT.
     ]{\resizebox{0.54\textwidth}{!}{%
        \begin{tikzpicture}

    \node[bsg_rectangle, minimum height=5cm, xshift=-0.5cm, font=\tiny, aspect=0.75,                                ] (n0 ) {User\\Logic};
    \node[bsg_rectangle, minimum width=0.5cm, below right=2.00cm and 2.25cm of n0 ] (n1a) {};
    \node[bsg_rectangle, minimum width=0.5cm,       right=0.5cm       of n1a] (n1b) {};
    \node[bsg_rectangle, minimum width=0.5cm,       right=1.25cm      of n1b] (n1c) {};
    \node[bsg_rectangle, minimum width=0.5cm,       above=0.3cm       of n1a] (n2a) {};
    \node[bsg_rectangle, minimum width=0.5cm,       right=0.5cm       of n2a] (n2b) {};
    \node[bsg_rectangle, minimum width=0.5cm,       right=1.25cm      of n2b] (n2c) {};
    \node[font=\tiny,                                 below left=0.30cm and 0.25cm of n1b] (t0 ){CSR synchronizers};
 
    \node[bsg_rectangle, minimum width=0.70cm, below right=1.15cm and 1.50cm of n0] (n3a) {};
    \node[bsg_rectangle, minimum width=0.70cm,       right=0.70cm       of n3a] (n3b) {};
    \node[bsg_rectangle, minimum width=0.70cm,       right=0.70cm       of n3b] (n3c) {};
    \node[bsg_rectangle, minimum width=0.70cm,       right=0.70cm       of n3c] (n3d) {};
    \node[bsg_rectangle, minimum width=0.25cm,       right=0.45cm       of n3d] (n3e) {};
    \node[bsg_rectangle, minimum width=0.25cm,       left=0.45cm        of n3a] (n3f) {};
    \node[bsg_rectangle, minimum width=0.70cm,       above=0.30cm        of n3a] (n4a) {};
    \node[bsg_rectangle, minimum width=0.70cm,       right=0.70cm       of n4a] (n4b) {};
    \node[bsg_rectangle, minimum width=0.70cm,       right=0.70cm       of n4b] (n4c) {};
    \node[bsg_rectangle, minimum width=0.70cm,       right=0.70cm       of n4c] (n4d) {};
    \node[bsg_rectangle, minimum width=0.25cm,       right=0.45cm       of n4d] (n4e) {};
    \node[bsg_rectangle, minimum width=0.25cm,       left=0.45cm        of n4a] (n4f) {};
    \node[font=\tiny,                                 below right=0.30cm and 0.43cm of n3b] (t1 ) {Asynchronous FIFOs};

\node [bsg_triangle , minimum width=0.4cm, rotate=270, above right=2.75cm and 0.20cm of n4e] (n6) {};
\node [bsg_triangle , minimum width=0.4cm, rotate=270, above right=2.75cm and 0.35cm of n4b] (n5) {};

\coordinate[below=0.53cm of n5] (b5);
\coordinate[below=1.00cm of n6] (b6);

\node [font=\tiny, above=0.40cm of n5] (t2) {DUT CLK};
\node [font=\tiny, above=0.40cm of n6] (t3) {\zynqps{} CLK};

\node [bsg_triangle , minimum width=0.4cm, rotate=270, above left=1.75cm and 0.35cm of n4f] (n7) {};
\coordinate[above left=0.50cm and 0.00cm of n7] (p0);
\node [font=\tiny, above=0.75cm of p0] (t4) {GATED CLK};

\node[bsg_cloud, font=\tiny, cloud puffs=18, xshift=-0.00cm, aspect=3.2, below=2.00cm of n5] (n8) {Gating FSM};
\node[bsg_cloud, font=\tiny, cloud puffs=12, xshift=0.05cm, aspect=1.5, below=2.00cm of n6] (n9) {P-Shell};
\node[bsg_cloud, font=\tiny, cloud puffs=15, xshift=0.00cm, aspect=2.3, below=1.05cm of n7] (n10) {Emulation};

\coordinate[right=0.70cm of n1c] (r1);
\coordinate[right=0.70cm of n2c] (r2);
\coordinate[right=0.60cm of n3e] (r3);
\coordinate[right=0.60cm of n4e] (r4);
\coordinate[left=1.50cm  of n1a] (l1);
\coordinate[left=1.50cm  of n2a] (l2);
\coordinate[left=0.50cm  of n7 ] (l7);
\coordinate[below=0.63cm of n7 ] (b7);
\coordinate[left=0.60cm  of n3f] (l3);
\coordinate[left=0.60cm  of n4f] (l4);
\coordinate[right=1.50cm of n7 ] (r7);
\coordinate[below=0.55cm of r7 ] (z0);

\path [->]
    (n5)     edge (n8)
    (n6)     edge (n9)
    (r1)     edge (n1c)
    (n2c)    edge (r2)
    (n3e)    edge (r3)
    (r4)     edge (n4e)
    (n1a)    edge (l1)
    (l2)     edge (n2a)
    (n1c)    edge (n1b)
    (n2b)    edge (n2c)
    (n7)     edge (n10)
    (p0)     edge (n7)
    (n4f)    edge (l4)
    (l3)     edge (n3f)
    (r7)     edge (n7)
    ;

\path [-]
    (b5)     edge (p0)
    (t2)     edge (n5)
    (t3)     edge (n6)
    (z0)     edge (r7)
    ;

\draw [loosely dashed] (1.50cm, -2.5cm) -- (1.50cm, 2.5cm);
\draw [loosely dashed] (3.60cm, -2.5cm) -- (3.60cm, 2.5cm);

\draw [dotted] (5.3cm,  1.00cm) -- (4.8cm,  2.25cm);
\draw [dotted] (5.3cm, -1.00cm) -- (4.8cm, -2.25cm);

\end{tikzpicture}
     }}
    \hspace{0.1cm}
    \subfloat[
        As DUT logic may be buggy during design, it is essential to not hang GP0, which could lead to \zynqps{} lockup. In \projectname{}, the \plshell{} prevents lockup regardless of DUT state, by lifting generic DUT interfaces to a set of nonblocking FIFO and read/write CSRs.
    ]{\resizebox{0.44\textwidth}{!}{%
        \begin{tikzpicture}
    \node[bsg_cloud,     minimum height=5cm, minimum width=2cm, cloud puffs=21                                   ] (n0)  {PL\\Logic};
    \node[bsg_rectangle, minimum height=5cm,     dashed,              right=                       of n0] (n1)  {AXI\\Regs};
    \node[bsg_rectangle, minimum height=5cm, minimum width=2.97cm, fill=none,         right=2.98cm of n0            ] (x0)  {};
    \node[above=2.75cm of x0] {\plshell{} Overlay};
    \node[bsg_rectangle, minimum size=2cm,                     right=2.5cm                  of n1] (n2)  {};
    \node[bsg_rectangle, minimum size=1.8cm,                   right=2.5cm                  of n1] (n3)  {PS};

    \node[bsg_rectangle, minimum width=1.3cm,                                      above left=1.8cm and 1.6cm of n1] (n4)    {FIFO 0};
    \node[bsg_rectangle, minimum width=1.3cm,                                      below=0.5cm of n4 ] (n5)  {CNT 0};
    \node[bsg_rectangle, minimum width=1.3cm,                                      below=0.6cm of n5 ] (n6)  {FIFO 1};
    \node[bsg_rectangle, minimum width=1.3cm,                                      below=0.5cm of n6 ] (n7)  {CNT 1};

    \node[bsg_rectangle, minimum width=1.3cm,                                      below=0.6cm of n7 ] (n8)  {CSR 0};
    \node[bsg_rectangle, minimum width=1.3cm,                                      below=0.55cm of n8 ] (n9)  {CSR 1};
    \node[bsg_rectangle, minimum width=1.3cm,                                      below=0.55cm of n9 ] (n10)  {CSR 2};
    \node[bsg_rectangle, minimum width=1.3cm,                                      below=0.55cm of n10] (n11)  {CSR 3};

    \coordinate[left=1.40cm of n4] (l4);
    \coordinate[left=1.40cm of n5] (l5);
    \coordinate[left=1.40cm of n6] (l6);
    \coordinate[left=1.40cm of n7] (l7);
    \coordinate[left=1.40cm of n8] (l8);
    \coordinate[left=1.40cm of n9] (l9);
    \coordinate[left=1.40cm of n10] (l10);
    \coordinate[left=1.40cm of n11] (l11);

    \coordinate[right=1.20cm of n4] (r4);
    \coordinate[right=1.20cm of n5] (r5);
    \coordinate[right=1.20cm of n6] (r6);
    \coordinate[right=1.20cm of n7] (r7);
    \coordinate[right=1.20cm of n8] (r8);
    \coordinate[right=1.20cm of n9] (r9);
    \coordinate[right=1.20cm of n10] (r10);
    \coordinate[right=1.20cm of n11] (r11);

\path [->]
    (n2)     edge[swap] node{GP0} (n1)
    
    (n4)     edge                 (l4)
    (l6)     edge                 (n6)

    (n8)     edge                 (l8)
    (n9)     edge                 (l9)
    (l10)    edge                 (n10)
    (l11)    edge                 (n11)
    
    (r4)     edge                 (n4)
    
    (n5)     edge                 (r5)

    (n6)     edge                 (r6)

    (n7)     edge                 (r7)

    (n8)     edge                 (r8)
    (r8)     edge                 (n8)

    (n9)     edge                 (r9)
    (r9)     edge                 (n9)

    (n10)    edge                 (r10)

    (n11)    edge                 (r11)
    ;

\draw [loosely dashed] (5.38cm, -2.5cm) -- (5.38cm, 2.5cm);

\node [below=2.7cm of n10] {};

\end{tikzpicture}
    }}
    \caption{The \projectname{} system provides system architects with full co-emulation capabilities through a simple C++ MMIO drivers, identically accessible from simulation, co-emulation or on deployed systems. Users parameterize the \plshell{} to for control or monitor execution, while the \zynqps{} runs any necessary software functional models.}
    \label{fig:02-architecture-hardware-architecture}
\end{figure*}
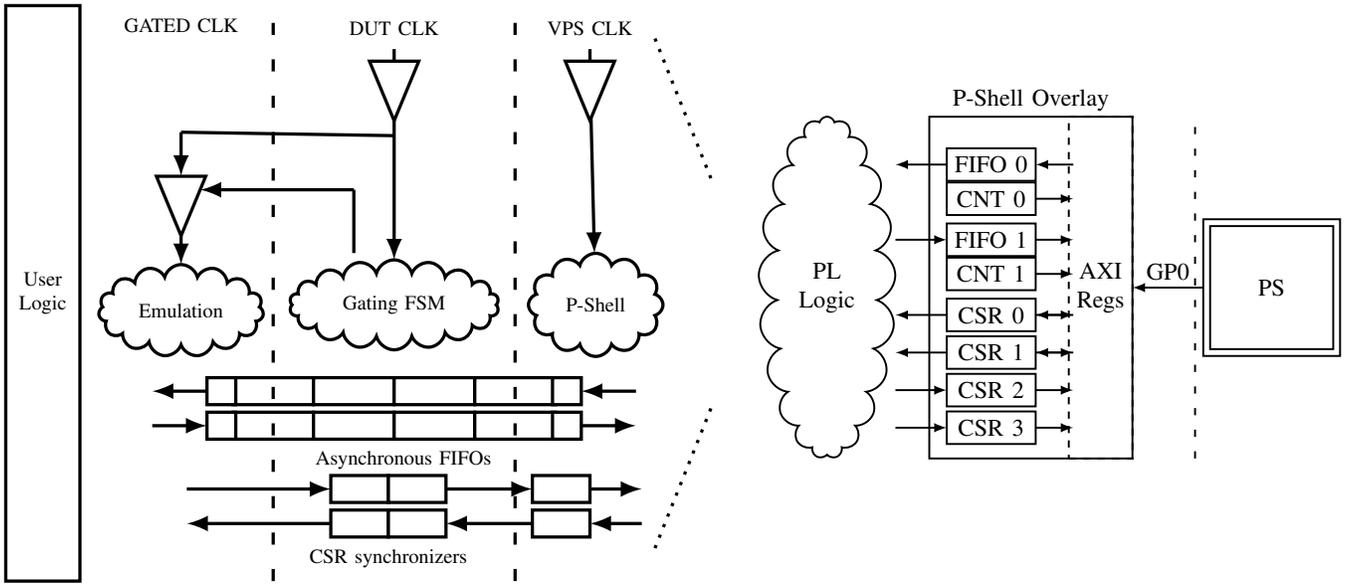

Zynq FPGA boards couple hardened ARM cores (\textit{\zynqps{}}) with a programmable fabric (\textit{PL}). Common peripherals such as USB, Ethernet, and DRAM are connected to the PS, while the PS communicates with the PL via hardened AXI~\cite{armaxi} interfaces. The \zynqps{} master ports are called \textit{GP} ports and cover a small address space. \zynqps{} client ports (\textit{HP}) are larger and higher performance, allowing the PL to indirectly access DRAM and peripherals.

\projectname{} leverages the Zynq architecture to decompose prototyping systems into these orthogonal functionalities. Bitstream generation can be done on any machine with a compatible Vivado version to the particular IP. From there, users can login to the PS over a standard Ethernet or UART connection, copy over the compiled bitstream and using the Pynq API, program the overlay and DUT on the PL.

\projectname{} provides an overlay (shown in Figure~\ref{fig:02-architecture-hardware-architecture}) that includes the \textit{\plshell{}}, the main interface between the host emulation and the DUT user logic. The \plshell{} provides a parameterizable array of input/output Control and Status Registers \textit{CSRs}, as well as an array of semi-blocking \textit{\nbfifo{}s}. An \nbfifo{} exposes blocking ready/valid~\cite{taylor2018basejump} interfaces to the PL side to support latency-insensitive interfaces, while the PS interacts with a non-blocking credit/valid interface to prevent system lockup. While non-blocking interfaces require multiple transactions for each read and write, they generally have little overall performance impact as the PS outpaces the PL during large system prototyping.

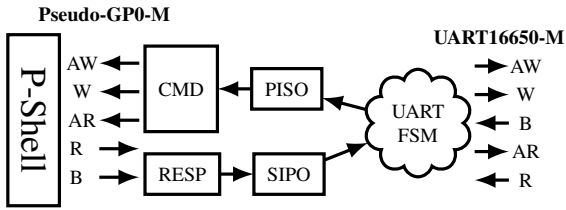
\begin{figure}
    \centering
    \resizebox{0.9\columnwidth}{!}{%
    \begin{tikzpicture}

\node[bsg_rectangle, minimum height=1.5cm] (n0) {\rotatebox{-90}{\small\plshell{}}};

\node[bsg_rectangle, minimum height=0.75cm, font=\tiny, below right= 0.10cm and 0.75cm of n0.north east] (n1) {CMD};
\node[bsg_rectangle, font=\tiny, above right= 0.10cm and 0.75cm of n0.south east] (n2) {RESP};

\node[bsg_rectangle, font=\tiny, right=0.30cm of n1.east] (n3) {PISO};
\node[bsg_rectangle, font=\tiny, right=0.30cm of n2.east] (n4) {SIPO};

\node[bsg_cloud, font=\tiny, inner sep=0.05cm, below right=0.30cm and 1.15cm of n3] (n5) {UART\\FSM};

\coordinate[above right= 0.50cm and 0.10cm of n5.east] (c1a); \coordinate[above right= 0.50cm and 0.4cm of n5.east] (c1b);
\coordinate[above right= 0.25cm and 0.10cm of n5.east] (c2a); \coordinate[above right= 0.25cm and 0.4cm of n5.east] (c2b);
\coordinate[above right= 0.00cm and 0.10cm of n5.east] (c3a); \coordinate[above right= 0.00cm and 0.4cm of n5.east] (c3b);
\coordinate[above right=-0.25cm and 0.10cm of n5.east] (c4a); \coordinate[above right=-0.25cm and 0.4cm of n5.east] (c4b);
\coordinate[above right=-0.50cm and 0.10cm of n5.east] (c5a); \coordinate[above right=-0.50cm and 0.4cm of n5.east] (c5b);

\draw[->] (c1a) -- (c1b) node[xshift=0.15cm, font=\tiny] {AW};
\draw[->] (c2a) -- (c2b) node[xshift=0.15cm, font=\tiny] {W};
\draw[->] (c3b) -- (c3a) node[xshift=0.45cm, font=\tiny] {B};
\draw[->] (c4a) -- (c4b) node[xshift=0.15cm, font=\tiny] {AR};
\draw[->] (c5b) -- (c5a) node[xshift=0.45cm, font=\tiny] {R};

\draw[->] (n2) -- (n4);
\draw[->] (n3) -- (n1);
\draw[->] (n4) -- (n5);
\draw[->] (n5) -- (n3);

\coordinate[above right= 0.50cm and 0.35cm of n0.east] (c6a);  \coordinate[above right= 0.50cm and 0.70cm of n0.east] (c6b);
\coordinate[above right= 0.25cm and 0.35cm of n0.east] (c7a);  \coordinate[above right= 0.25cm and 0.70cm of n0.east] (c7b);
\coordinate[above right= 0.00cm and 0.35cm of n0.east] (c8a);  \coordinate[above right= 0.00cm and 0.70cm of n0.east] (c8b);
\coordinate[above right=-0.25cm and 0.35cm of n0.east] (c9a);  \coordinate[above right=-0.25cm and 0.70cm of n0.east] (c9b);
\coordinate[above right=-0.50cm and 0.35cm of n0.east] (c10a); \coordinate[above right=-0.50cm and 0.70cm of n0.east] (c10b);

\draw[->] (c6b)  node[xshift=-.50cm, font=\tiny] {AW} -- (c6a);
\draw[->] (c7b)  node[xshift=-.50cm, font=\tiny] {W}  -- (c7a);
\draw[->] (c8b)  node[xshift=-.50cm, font=\tiny] {AR} -- (c8a);
\draw[->] (c9a)  node[xshift=-.20cm, font=\tiny] {R}  -- (c9b);
\draw[->] (c10a) node[xshift=-.20cm, font=\tiny] {B}  -- (c10b);

\node[above right=0.00cm and -0.30cm of n0.north east] {\textbf{\tiny{Pseudo-GP0-M}}};
\node[above right=0.25cm and -0.30cm of n5.north east] {\textbf{\tiny{UART16650-M}}};

\end{tikzpicture}
    }
    \caption{When prototyping on non-Zynq FPGAs, \projectname{} transparently tunnels requests into a "pseudo-GP0" accessing the \plshell{}. For instance, GP0 writes are deserialized from UART RX while GP0 reads are deserialized from RX and then reserialized into UART TX.}
    \label{fig:02-architecture-uart-bridge}
\end{figure}

While \projectname{} fits seamlessly into the Zynq PS-PL paradigm, there are many FPGA architectures which lack hardened CPU cores. For these boards, \projectname{} provides hardware bridges (shown in Figure~\ref{fig:02-architecture-uart-bridge}) which can convert C++ \plshell{} requests to a pseudo-GP master bus via a transparent software translation. We refer to this combination of C++ co-simulation code, host transport layer and GP master bus as a \textit{\zynqps{}}. The \zynqps{} abstraction supports the flexibility of arbitrary MMIO interaction with the DUT while switching out transport layers optimized for the specific execution environment, all while maintaining lockup safety and reasonable performance.

When prototyping ASICs, the DUT clock is often limited by poor mapping of standard cells to FPGA primitives~\cite{wong2017superscalar}, limiting the overall emulation performance. While some efficiency may be regained by explicit manual remapping of problematic primitives (CAMs, large muxes, heavily retimed modules), this duplicates design efforts and forces dependencies between FPGA and ASIC teams. Unfortunately, even the best mapping efforts cannot solve this problem for aggressive submicron designs. To alleviate performance bottlenecks and decouple the prototype and emulation, asynchronous FIFOs and CSR synchronizers bridge the \zynqps{} and DUT clock domains. This decoupling allows the \zynqps{} to run ahead of PL and averts complex emulation models slowing the DUT execution.

\subsection{Emulation Layer}

\begin{figure*}
    \centering
    \resizebox{1.00\textwidth}{!}{%
    \newsavebox\codeboxa
\newsavebox\codeboxb
\newsavebox\codeboxc
\newsavebox\codeboxd
\newsavebox\codeboxe

\begin{lrbox}{\codeboxa}
\begin{lstlisting}[language=C++]
#include <pshell.hpp>

// Demonstrate common interfacing
//   with loopback DUT through P-Shell
void ps_main()
{
    // Initialize P-Shell
    pshell_t *pl = new pshell_t();

    // Write 0xbeef to CSR A
    pl->write(SHELL_CSR_A, 0xbeef);
    // Read back that 0xbeef
    int val1 = pl->read(SHELL_CSR_A);

    // Wait for space in FIFO X
    while (!pl->read(SHELL_FIFO_X_CNT));
    // Write 0xcafe to FIFO X
    pl->write(SHELL_FIFO_X, 0xcafe);

    // Wait for response in FIFO Y
    while (!pl->read(SHELL_FIFO_Y_CNT));
    // Read 0xcafe from FIFO Y
    int val2 = pl->read(SHELL_FIFO_Y);
}
\end{lstlisting}
\end{lrbox}

\begin{lrbox}{\codeboxb}
\begin{lstlisting}[language=C++]
// Do ARM PS read
int axi_read(int addr) {
    return *((int *)gp0_ptr+addr);
}
// Do ARM PS write
void axi_write(int addr, int data) {
    *((int *)gp0_ptr+addr) = data;
}
\end{lstlisting}
\end{lrbox}

\begin{lrbox}{\codeboxc}
\begin{lstlisting}[language=C++]
// Do DPI read
int axi_read(int addr) {
    gp0_arvalid = 1;
    do { yield(); } while(!gp0_arready);
    gp0_arvalid = 0; 
    int val = gp0_rdata;
    do { yield(); } while(!gp0_rvalid);
    return val;
}
// Do DPI write
void axi_write(int addr, int data) {
    gp0_awvalid = 1; gp0_wvalid = 1;
    do {
        awdone = gp0_awready;
        gp0_awvalid = !awdone;
        wdone = gp0_wready;
        gp0_wvalid = !wdone;
        yield();
    } while (!awdone || !wdone);
    gp0_bready = 1;
    do { yield(); } while(!gp0_bvalid);
}
\end{lstlisting}
\end{lrbox}

\begin{lrbox}{\codeboxd}
\begin{lstlisting}[language=C++]
// Initilize PS MMIO
void init() {
    gp0_ptr = mmap(GP0_PADDR);
    dram_ptr = cma_alloc(PL_DRAM_SIZE);
}
\end{lstlisting}
\end{lrbox}

\begin{lrbox}{\codeboxe}
\begin{lstlisting}[language=C++]
// Initialize DPI GPIO C
void init() {
    gp0_arvalid = new gpio_dpi();      
    gp0_awvalid = new gpio_dpi();
    dram_ptr = malloc(PL_DRAM_SIZE);
    ...
}
\end{lstlisting}
\end{lrbox}

\begin{tikzpicture}

\node[bsg_rectangle,                                                ] (n0) {\resizebox{8.0cm}{!}{\usebox\codeboxa}};
\node[bsg_rectangle,  below left=-0.75cm and 0.50cm of n0.north west] (n1) {\resizebox{6.5cm}{!}{\usebox\codeboxc}};
\node[bsg_rectangle,  above left=-0.75cm and 0.50cm of n0.south west] (n2) {\resizebox{6.5cm}{!}{\usebox\codeboxb}};

\node[bsg_rectangle,   below left=0.25cm and 0.5cm     of n1.north west ] (n3a) {DPI GPIO (Verilog)};
\node[bsg_rectangle,   below=0.70cm of n3a] (n3b) {DPI GPIO (Verilog)};
\node[bsg_rectangle,   below=0.70cm of n3b] (n3c) {DPI GPIO (Verilog)};
\node[bsg_rectangle,   below=0.70cm of n3c] (n3d) {DPI GPIO (Verilog)};
\node[bsg_rectangle,   below=0.70cm of n3d] (n3e) {DPI GPIO (Verilog)};
\node[bsg_rectangle,   below=0.70cm of n3e] (n3f) {DPI GPIO (Verilog)};
\node[bsg_rectangle,   below=0.70cm of n3f] (n3g) {DPI GPIO (Verilog)};
\node[bsg_rectangle,   below=0.70cm of n3g] (n3h) {DPI GPIO (Verilog)};
\node[bsg_rectangle,   below=0.70cm of n3h] (n3i) {DPI GPIO (Verilog)};
\node[                 below=0.70cm of n3i] (n3j) {...};
\node[bsg_rectangle,   below=0.70cm of n3j] (n3k) {DPI GPIO (Verilog)};

\node[bsg_rectangle, minimum height=3cm, minimum width=3cm, left=0.5cm of n2.west] (n4) {};
\node[bsg_rectangle, minimum height=2.8cm, minimum width=2.8cm, left=0.6cm of n2.west] {\zynqps{}};
\node[bsg_rectangle, minimum height=2cm, minimum width=1cm, left=0.35cm of n4.west] (n7) {\rotatebox{270}{P-Shell}};
\node[bsg_cloud, minimum width=3.5cm, aspect=1, left=0.35cm of n7.west] (n8) {PL Logic};
\draw (n7) edge (n8);
\draw (n4) edge (n7);

\node[bsg_rectangle, inner sep=0cm, minimum height=8cm, above left=4cm and 4.5cm of n4.south west] (t0) {};
\node[bsg_rectangle, inner sep=0cm, minimum height=8cm, above left=4cm and 2.5cm of n4.south west] (t1) {};
\node[bsg_rectangle, inner sep=0cm, minimum height=8cm, above left=4cm and 0.5cm of n4.south west] (t2) {};

\node[above=0.0cm of t0.north] {\large Simulator};
\node[above=0.1cm of t1.north] (t3) {\large PS};
\node[above=0.0cm of t2.north] {\large Peripheral};
\node[above=0.0cm of t3.north] {\Large\textbf{Coroutines}};

\node[bsg_rectangle, above=0.75cm of n0.north] (n5) {\resizebox{8.0cm}{!}{\usebox\codeboxe}};
\node[bsg_rectangle, below=0.75cm of n0.south] (n6) {\resizebox{8.0cm}{!}{\usebox\codeboxd}};

\coordinate[left=20cm of n5] (c5);
\coordinate[left=20cm of n6] (c6);
\coordinate[left=20cm of n0, yshift=-2.25cm] (c0);

\draw (n0.west) edge[yshift=-2.30cm, dashed, ultra thick] (c0); 

\path [->]
    (n5.west) edge[ultra thick, bend right=5] node[yshift=1.00cm,xshift=0.25cm] {\huge\textbf{RTL Co-Simulation}} (c5)
    (n6.west) edge[ultra thick, swap, bend left=5]  node[yshift=0.25cm] {\huge\textbf{Synthesized Co-Emulation}} (c6)
    ;

\coordinate[below=0.5cm of t0.north] (s0a);
\coordinate[below=0.5cm of t1.north] (s0b);
\coordinate[below=0.5cm of t2.north] (s0c);
\coordinate[below=1.5cm of t0.north] (s1a); \node[left=0.10cm of s1a] {tick};
\coordinate[below=1.5cm of t1.north] (s1b);
\coordinate[below=1.5cm of t2.north] (s1c);
\coordinate[below=2.5cm of t0.north] (s2a);
\coordinate[below=2.5cm of t1.north] (s2b);
\coordinate[below=2.5cm of t2.north] (s2c);
\coordinate[below=3.5cm of t0.north] (s3a);  \node[left=0.10cm of s3a] {tick};
\coordinate[below=3.5cm of t1.north] (s3b);
\coordinate[below=3.5cm of t2.north] (s3c);
\coordinate[below=4.5cm of t0.north] (s4a);
\coordinate[below=4.5cm of t1.north] (s4b);
\coordinate[below=4.5cm of t2.north] (s4c);
\coordinate[below=5.5cm of t0.north] (s5a);  \node[left=0.10cm of s5a] {tick};
\coordinate[below=5.5cm of t1.north] (s5b);
\coordinate[below=5.5cm of t2.north] (s5c);
\coordinate[below=6.5cm of t0.north] (s6a);
\coordinate[below=6.5cm of t1.north] (s6b);
\coordinate[below=6.5cm of t2.north] (s6c);
\coordinate[below=7.5cm of t0.north] (s7a);  \node[left=0.10cm of s7a] {tick};
\coordinate[below=7.5cm of t1.north] (s7b);
\coordinate[below=7.5cm of t2.north] (s7c);

\path [->]
    (s0a) edge node[font=\small, rotate=0,   xshift=0.25cm] {req start} (s0b)
    (s0b) edge[swap] node[font=\small, rotate=30,   xshift=0.25cm] {yield} (s1a)
    (s1a) edge node[font=\small, rotate=0,   xshift=0.30cm] {req cont.} (s1b)
    (s1b) edge node[font=\small, rotate=0, xshift=0.10cm] {has req?} (s1c)
    (s1c) edge[swap] node[font=\small, rotate=30, xshift=0.30cm] {req start} (s2b)
    (s2b) edge[swap] node[font=\small, rotate=30, xshift=0.25cm] {yield} (s3a)
    (s3a) edge[bend left] node[font=\small] {wait} (s5a)
    (s5a) edge node[font=\small, xshift=0.00cm] {req end} (s5b)
    (s5b) edge node[font=\small, rotate=0, xshift=0.10cm] {has req?} (s5c)
    (s5c) edge[swap] node[font=\small, rotate=30, xshift=0.35cm] {req end} (s6b)
    (s6b) edge[swap] node[font=\small, rotate=30, xshift=0.25cm] {yield} (s7a)
    ;

\end{tikzpicture}
    }
    \caption{\projectname{} enables designs to run identical C++ code on the \zynqps{} of a Zynq ARM core, over a UART bridge or in vendor-agnostic simulation. Instead of relying on Verilog tasks to interact with the DUT, \projectname{} exposes pins on the \plshell{} through a DPI-C interface. The result is fine-grained control over DUT execution, enabling software flow-control and thorough verification. As multithreading is disallowed by many commerical simulators, C++ coroutines are used to co-simulate the DUT with blocking transactions such as AXI requests, providing parallelism and deadlock avoidance.}
    \label{fig:02-architecture-axi-abstraction}
\end{figure*}

During deep performance profiling, the \zynqps{} may need to process monitoring information or system-call emulation every DUT cycle while it is also handling context switching, network bridging or other asynchronous processing. If the \zynqps{} is not ready to accept a new packet and an asynchronous FIFO fills, either the FIFO must backpressure such that cycle-accuracy is lost, or the packet is dropped. Most systems using latency-insensitive I/O constructs use ready/valid handshakes to pause the DUT operation upon backpressure. However, doing so perturbs the system and eliminates cycle-accuracy, making the emulation non-reproducible. Figure~\ref{fig:02-architecture-axi-abstraction} shows an abstraction of the \zynqps{} responsibilities.

Another approach to avoid degradation is to run a Real-Time Operating System (RTOS) on the \zynqps{}. However hard real-time guarantees are difficult to meet, restricting maximum performance; and proofs would need to be rewritten for each DUT interface, slowing iteration time. During profiling the information bandwidth needed varies dramatically based on the specific performance aspect being monitored. These guarantees will need to be retuned for each monitoring mode, as well as relaxed for running in a simulation mode that has dramatically different (wall-clock) timing characteristics. In addition, compiling arbitrary programs is much more difficult on a specialized RTOS compared to a full POSIX operation system.

\projectname{} leverages the \zynqps{}-DUT asynchrony to implement cycle-accurate emulation by gating the DUT clock upon interfering backpressure. Once gated, the asynchronous FIFOs are drained and execution can safely resume. This approach masks non-determinism in the \zynqps{}, which may be running a full PetaLinux~\cite{petalinux} operating system. Clock gating in the \plshell{} means that both \zynqps{} software and DUT logic can be completely unaware of the other side of the interface, operating in an ideal environment. Clearly defined boundaries between \zynqps{} and DUT domains simplify necessary timing constraints during synthesis and standardized, validated asynchronous primitives shield users from the subtle gotchas of multi-clock systems.

On the other hand, modelling exact I/O timing is an essential functionality in a Scale-Down system. In \projectname{}, hardware model timers exist in the DUT clock domain, but timing information is stored in the \zynqps{} program where it is exchanged via a simple handshake. For instance to prototype a system with cutting-edge HBM DRAM, the DUT may emit a DRAM request which causes DUT execution to stop. The \zynqps{} receives the request, calculates the predicted timing of the specific HBM model, and programs the expected timing through \plshell{} CSRs. The DUT clock then resume, waiting for the DRAM request to return but executing any other parallel tasks. If the DRAM request returns before the hardware model timer finishes, it will be paused until the correct cycle. If the hardware model timer expires before the DRAM request returns, the DUT will return to a gated state. This event-driven co-emulation maintains cycle-accuracy while ensuring there are minimal wasted cycles.

\section{\projectname{} Decomposition}

\begin{figure}
    \centering
    \begin{tikzpicture}

\node[bsg_rectangle, minimum width=1.50cm, minimum height=1.50cm] (n0) {};
\node[bsg_rectangle, minimum width=0.3cm, minimum height=1cm, right=1.50cm of n0.east] (n0shell) {\rotatebox{90}{\plshell{}}};
\node[above=0.00cm of n0.north] (n0l) {Multicore};

\node[bsg_rectangle, minimum width=0.60cm, minimum height=0.60cm, below right=0.10cm and 0.10cm of n0.north west] (n0a) {C};
\node[bsg_rectangle, minimum width=0.60cm, minimum height=0.60cm, below left=0.10cm and 0.10cm of n0.north east] (n0b) {C};
\node[bsg_rectangle, minimum width=0.60cm, minimum height=0.60cm, above right=0.10cm and 0.10cm of n0.south west] (n0c) {C};
\node[bsg_rectangle, minimum width=0.60cm, minimum height=0.60cm, above left=0.10cm and 0.10cm of n0.south east] (n0d) {C};

\node[bsg_rectangle, minimum width=1.50cm, minimum height=1.50cm, below=0.50cm of n0.south] (n1) {};
\node[bsg_rectangle, minimum width=0.3cm, minimum height=1cm, right=1.50cm of n1.east] (n1shell) {\rotatebox{90}{\plshell{}}};
\node[above=0.00cm of n1.north] (n1l) {Core};

\node[bsg_rectangle, minimum width=0.30cm, minimum height=1.30cm, above right=0.10cm and 0.10cm of n1.south west] (n1a) {\small{FE}};
\node[bsg_rectangle, minimum width=0.30cm, minimum height=1.30cm, above left=0.10cm and 0.10cm of n1.south east] (n1b) {\small{BE}};

\node[bsg_rectangle, minimum width=1.50cm, minimum height=1.50cm, right=1.00cm of n0shell.east] (n2) {};
\node[bsg_rectangle, minimum width=0.3cm, minimum height=1cm, right=1.50cm of n2.east] (n2shell) {\rotatebox{90}{\plshell{}}};
\node[above=0.00cm of n2.north] (n2l) {Front End};

\node[bsg_rectangle, minimum width=0.50cm, minimum height=0.50cm, above right=0.10cm and 0.10cm of n2.south west] (n2a) {\rotatebox{90}{\tiny{BTB}}};
\node[bsg_rectangle, minimum width=0.50cm, minimum height=0.50cm, above=0.10cm of n2a.north] (n2b) {\rotatebox{90}{\tiny{BHT}}};
\node[bsg_rectangle, minimum width=0.60cm, minimum height=1.30cm, above left=0.10cm and 0.10cm of n2.south east] (n2c) {\rotatebox{90}{\footnotesize{ICACHE}}};

\node[bsg_rectangle, minimum width=1.50cm, minimum height=1.50cm, right=1.00cm of n1shell.east] (n3) {};
\node[bsg_rectangle, minimum width=0.3cm, minimum height=1cm, right=1.50cm of n3.east] (n3shell) {\rotatebox{90}{\plshell{}}};
\node[above=0.00cm of n3.north] (n3l) {I-Cache};

\node[bsg_rectangle, minimum width=0.30cm, minimum height=1.30cm, above right=0.10cm and 0.10cm of n3.south west] (n3a) {\rotatebox{90}{\scriptsize{DATA}}};
\node[bsg_rectangle, minimum width=0.30cm, minimum height=1.30cm, right=0.05cm of n3a.east] (n3b) {\rotatebox{90}{\scriptsize{TAG}}};
\node[bsg_rectangle, minimum width=0.30cm, minimum height=1.30cm, right=0.05cm of n3b.east] (n3c) {\rotatebox{90}{\scriptsize{STAT}}};

\path [<->]
    (n0) edge[<->,transform canvas={yshift=0.45cm}] node[font=\footnotesize, rotate=0] {AXI} (n0shell)
    (n0) edge[<-,transform canvas={yshift=0.00cm}] node[font=\footnotesize, rotate=0] {CHIP ID} (n0shell)
    (n0) edge[->,transform canvas={yshift=-0.45cm}] node[font=\footnotesize, rotate=0] {STATS} (n0shell)

    (n1) edge[<->,transform canvas={yshift=0.45cm}] node[font=\footnotesize, rotate=0] {NOC} (n1shell)
    (n1) edge[<-,transform canvas={yshift=0.00cm}] node[font=\footnotesize, rotate=0] {CORE ID} (n1shell)
    (n1) edge[->,transform canvas={yshift=-0.45cm}] node[font=\footnotesize, rotate=0] {DMA} (n1shell)

    (n2) edge[->,transform canvas={yshift=0.45cm}] node[font=\footnotesize, rotate=0] {PC} (n2shell)
    (n2) edge[->,transform canvas={yshift=0.00cm}] node[font=\footnotesize, rotate=0] {INSTR} (n2shell)
    (n2) edge[<-,transform canvas={yshift=-0.45cm}] node[font=\footnotesize, rotate=0] {IRQ} (n2shell)

    (n3) edge[<-,transform canvas={yshift=0.45cm}] node[font=\footnotesize, rotate=0] {PC} (n3shell)
    (n3) edge[->,transform canvas={yshift=0.00cm}] node[font=\footnotesize, rotate=0] {INSTR} (n3shell)
    (n3) edge[<->,transform canvas={yshift=-0.45cm}] node[font=\footnotesize, rotate=0] {FILL} (n3shell)
    ;
   

\end{tikzpicture}
    \caption{Decomposition into smaller components leverage smaller FPGAs to verify modules in parallel as well as hyperfocus on specific performance or verification targets. In the above decomposition, verification engineers can tune their analysis for either a full multicore, a single core, the front end or just the instruction cache. Because the \plshell{} supports both unidirectional and latency-insensitive links, any hardware interface can be exposed. Rather than cumbersome hardened RTL models, \zynqps{} software mimics the interface timings.}
    \label{fig:02-architecture-partial}
\end{figure}
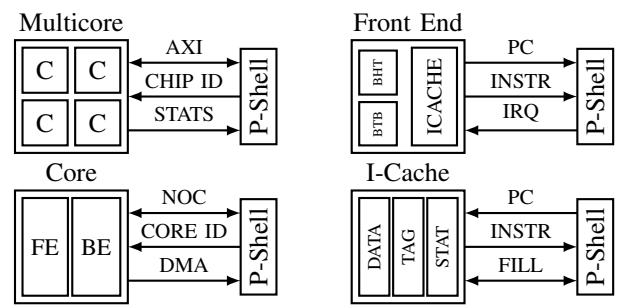

A key element of \scaledown{} is shrinking the design size while maintaining the integrity of inputs and outputs of the system. As illustrated in Figure~\ref{fig:02-architecture-partial}, even for a single design hierarchy there are various decomposition strategies. The best strategy will depend on the size of the design, verification team and FPGA supply. At later stages in the verification cycle it may be advantageous to hyperfocus on smaller modules, attempting to expose subtle optimizations that may be clouded by full-system effects. Small, cost-effective FPGAs are especially advantageous at this stage. Firstly, smaller designs will benefit from shorter compilation times and faster emulation speeds, leading to quicker overall turnaround time. Secondly, when running a large number of tests (for example running a benchmark suite on a processor design), each independent run requires system resources such as a DRAM storage and bandwidth to support it. Allocating a cheap board for each test scales testing throughput linearly, while investing in larger FPGAs quickly becomes cost prohibitive. 

\section{Case Studies: \projectname{} in the Wild}

We first developed \projectname{} for undergraduate/graduate architecture class performing full-stack analysis and optimizations to an open-source Linux-capable RISC-V multicore. Our main goals were:
\begin{enumerate}
    \item Provide a cheap, flexible platform for designing and analyzing microarchitectural modifications to the core.
    \item Avoid supporting all laptop to FPGA mappings by standardizing the host system (the Zynq PS).
    \item Synchronize co-simulation and co-emulation execution to minimize FPGA debugging.
    \item Make the platform robust to fatal RTL bugs by construction, impossible to hang the host system.
\end{enumerate}

However, we quickly realized that the abstract capability of interacting out-of-band with arbitrary devices was applicable in a large number of diverse settings. In this section we describe a few use-cases that we have found for \projectname{} since its inception.

\subsection{A Scale-Down \farm{} Cluster}

We built the first \projectname{} on TUL Z2~\cite{pynqz2}, inexpensive educational boards available at an academic discount. Z2 boards are out-of-box compatible with the open-source Xilinx Pynq~\cite{pynq} SDK, providing a Python-based interface for bitstream programming, peripheral management and \zynqps{} configuration, among many other convenience features. Because the Pynq software makes interaction with the Z2 boards so convenient, students can buy and develop on their own device. Most FPGA development boards, including the Z2, feature a watchdog timer which forces a reset upon hanging the board. In academia, this feature is invaluable to ensure that inexperienced students can always access their board.

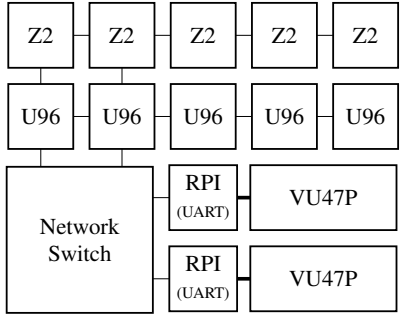
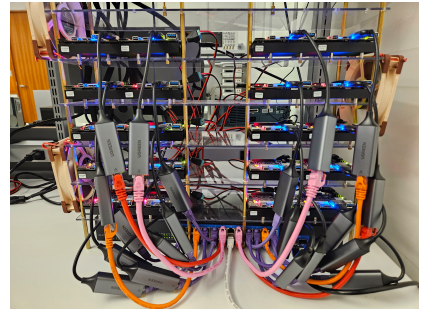
\begin{figure*}
    \centering
    \subfloat[
         \projectname{} clusters connect to a standard network switch to enable remote connections. While homogeneous clusters of Pynq Boards is the lowest maintenance options, some labs may be restricted to non-Zynq FPGAs and use small controllers such as Raspberry Pi to bridge to a \zynqps{} interface.
     ]{\makebox[0.49\textwidth]{\resizebox{0.3\textwidth}{!}{%
        \begin{tikzpicture}

\node[bsg_rectangle, minimum height=1cm, minimum width=1cm] (n1) {Z2};
\node[bsg_rectangle, minimum height=1cm, minimum width=1cm, left=1.25cm  of n1] (n2) {Z2};
\node[bsg_rectangle, minimum height=1cm, minimum width=1cm, left=1.25cm  of n2] (n3) {Z2};
\node[bsg_rectangle, minimum height=1cm, minimum width=1cm, right=1.25cm of n1] (n4) {Z2};
\node[bsg_rectangle, minimum height=1cm, minimum width=1cm, right=1.25cm of n4] (n5) {Z2};

\node[bsg_rectangle, minimum height=1cm, minimum width=1cm, below=1.25cm of n1] (n6)  {U96};
\node[bsg_rectangle, minimum height=1cm, minimum width=1cm, left=1.25cm  of n6] (n7)  {U96};
\node[bsg_rectangle, minimum height=1cm, minimum width=1cm, left=1.25cm  of n7] (n8)  {U96};
\node[bsg_rectangle, minimum height=1cm, minimum width=1cm, right=1.25cm of n6] (n9)  {U96};
\node[bsg_rectangle, minimum height=1cm, minimum width=1cm, right=1.25cm of n9] (n10) {U96};

\node[bsg_rectangle, minimum height=1cm, minimum width=2.25cm, below right=1.25cm and 0.60cm of n9 ] (n11) {VU47P};
\node[bsg_rectangle, minimum height=1cm, minimum width=2.25cm, below=1.25cm                  of n11] (n12) {VU47P};

\node[bsg_rectangle, minimum height=1cm, minimum width=1cm, left=1.85cm of n11] (n13) {RPI\\\scriptsize{(UART})};
\node[bsg_rectangle, minimum height=1cm, minimum width=1cm, left=1.85cm of n12] (n14) {RPI\\\scriptsize{(UART})};

\node[bsg_rectangle, minimum height=2.25cm, minimum width=2.25cm, below right=1.90cm and 0.60cm of n8] {Network\\Switch};

\coordinate[left=1cm of n13] (l13);
\coordinate[left=1cm of n14] (l14);
\coordinate[below=1cm of n7] (b7);
\coordinate[below=1cm of n8] (b8);

\begin{scope}[on background layer]
\path [-]
    (n3)     edge              (n5)
    (n8)     edge              (n10)
    (n13)    edge[ultra thick] (n11)
    (n14)    edge[ultra thick] (n12)
    (n13)    edge              (l13)
    (n14)    edge              (l14)
    (n2)     edge              (b7)
    (n3)     edge              (b8)
    ;
\end{scope}

\end{tikzpicture}
     }}}
    \hspace{0.10cm}
    \subfloat[
        A twenty-server Ultra96v2 cluster. Students can time-share boards for parallel builds and serialized, private experiments. By connecting the cluster to a network switch, students are able to work fully remotely, important during events such as the COVID-19 pandemic.
    ]{\makebox[0.49\textwidth]{\resizebox{0.3\textwidth}{!}{%
        \includegraphics{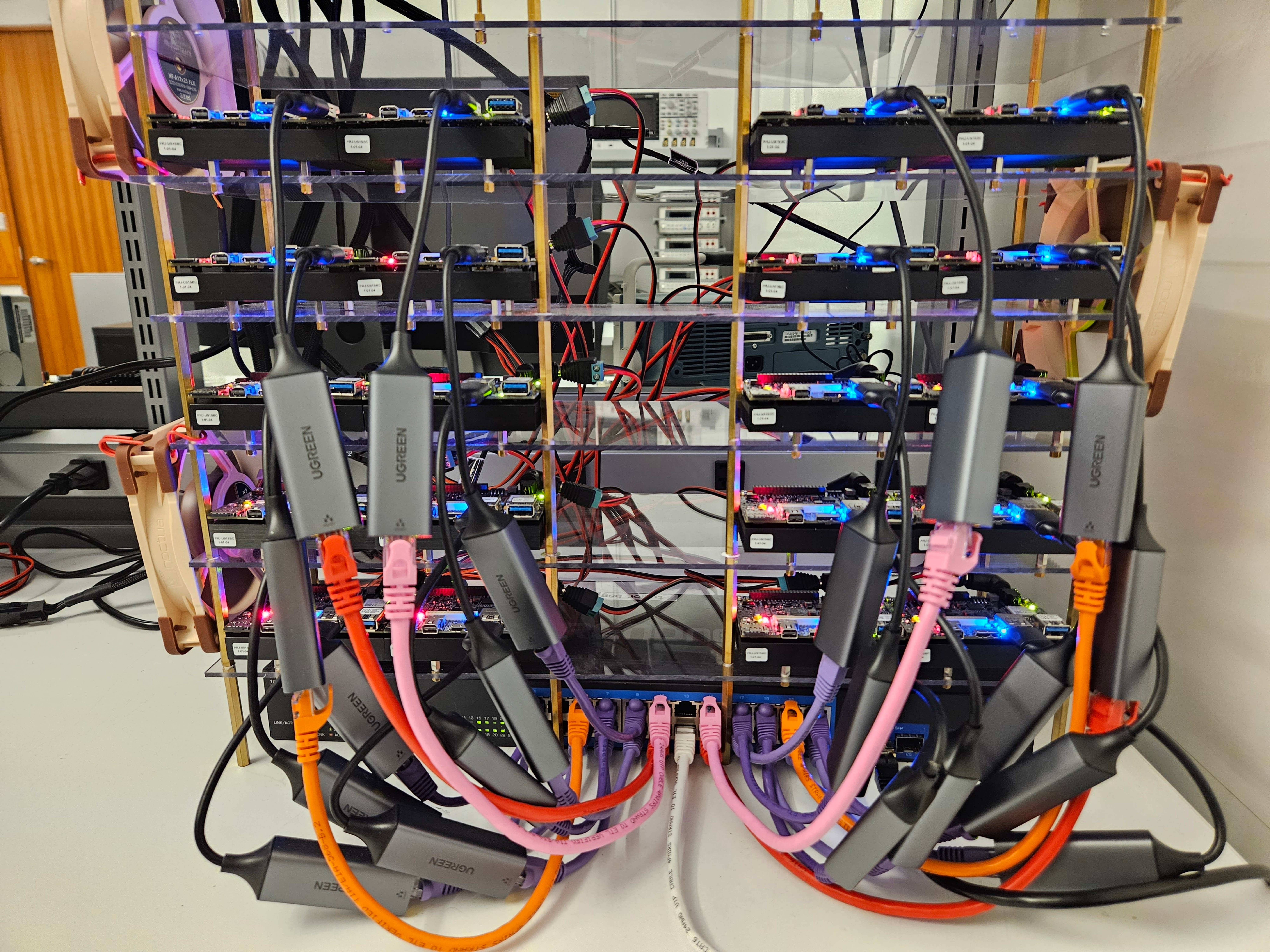}
    }}}
    \caption{
        \farm{} can be configured for a variety of Pareto frontiers along cost, capacity and design parallelism. For a design space exploration of heterogenous components, a fleet of small FPGAs may minimize build times, whereas for a suite of long-running benchmarks, medium-sized FPGAs may be able to complete overnight regressions on a full system.
        }
    \label{fig:02-farm-unified}
\end{figure*}

Needing a more structured approach to coherently integrate a large number of boards, we designed the (\textit{\farm{}}), a scalable cluster of network-attached FPGAs running \projectname{}. All \projectname{} data is stored on a host system and shared with the boards through a network-based distributed file-system. For parallel development, team members log into each board to independently program and run experiments. Bulk regression can be run from standard job-scheduling software. The setup shown in Figure~\ref{fig:02-farm-unified}(b) is built with commodity components: USB-Ethernet controllers, a network switch, and hand-cut plexiglass shelves. Comprising 20 Ultra96v2 boards, this \farm{} cost around \$4500 and supports multiple projects and Continuous Integration (\textit{CI}) runners for a modestly sized research group. Based on Table~\ref{tab-01-introduction:comparison}, we estimate that a \farm{} outperforms in TCO after less than a full year of usage.

Maintaining a \farm{} is typically as simple as ensuring the central network switch is remotely accessible. If the watchdog timers are properly configured, any temporary glitch with the board has a fail-safe backup and connections can generally be restored after reboot. For further robustness, a remote network-attached reset switch (or Raspberry Pi~\cite{raspberrypi}) removes the need to physically reset the system even upon unlikely watchdog failures. A centralized job-scheduler can dynamically prevent interference between users and regression jobs.

As shown in Figure~\ref{fig:02-farm-unified}(a), \projectname{} easily supports heterogeneous \farm{}s as there are only two classes of network interface to maintain. Standard Zynq boards connect directly via Ethernet while non-Zynq parts tunnel through a network-attached UART-capable device such as a Raspberry Pi. In this way, clusters can simultaneously service a wide range of IP blocks that each may leverage specific board features. A diverse setup is ideal for a large continuous integration server as generic jobs can be assigned to minimally-sized boards, reducing regression time and improving energy efficiency. 

To ensure the correctness of the \blackparrot{} execution and the observed performance data, user needs to be able to verify the correct execution of a benchmark during its runtime. For example, \projectname{}'s infrastructure can also be used to extract instruction commit information of a RISC-V core and cross-verify it with Dromajo~\cite{dromajo} abstract software-based golden model of DUT in \zynqps{}. Similarly, on each instruction commit and register write, corresponding information is written to asynchronous FIFOs that can gate the DUT clock while \zynqps{} drains and verifies the data. Using this feature, \projectname{} can be integrated into the CI as part of the chip development cycle. On a major change, FPGAs can be used to accelerate design verification using longer benchmarks that are impractical for RTL simulation.

\subsection{Microarchitectural Optimization: \catchup{} ALU}

In addition to modelling interface timings, the \plshell{} can be used for verification of \blackparrot{} logic by 



extracting commit information and cross-verifying it with Dromajo~\cite{dromajo}, an abstract software-based golden model. Upon each commit PC, instruction metadata, and writeback information are written to asynchronous FIFOs. \zynqps{} backpressure gates the DUT clock as \zynqps{} drains the commits. With cycle-accurate co-emulation, \projectname{} can be integrated into the CI as part of the chip development cycle.

Because \projectname{} is able to maintain cycle-accuracy with arbitrary bandwidth instrumentation, it enables deep insight into subtle microarchitectural bottlenecks. Previous works have proposed sampling-based architectures that accurately detect long latency stalls such as page table walks and L1 cache misses, but cannot diagnose ultra fine-grained stall sources such as irregular dependency bubbles. Section~\ref{sec:10-case-sampling} more thoroughly explores the trade-off between emulation speed and attribution accuracy, demonstrating the need for ultra fine-grained sampling to detect certain types of microarchitectural bottlenecks.

\begin{figure}
    \centering
    \includegraphics[width=\columnwidth]{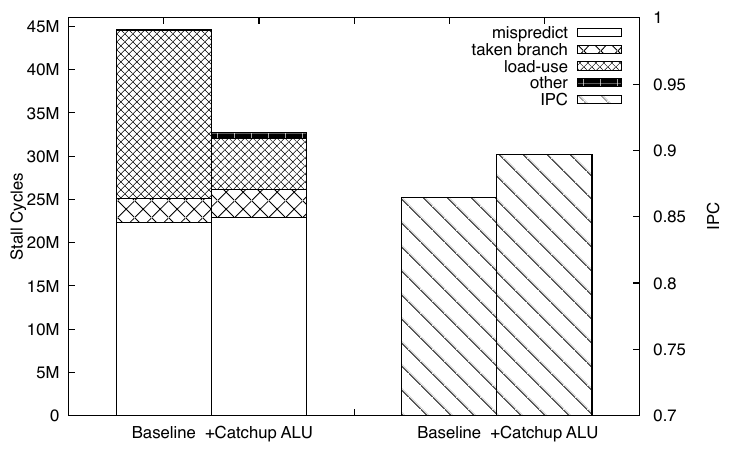}
    \caption{After basic core optimization, remaining low latency stalls (1-5 cycles) are difficult to detect via coarse-grained sampling. Tailored event counters can identify problematic categories, but lose PC association during aggregation. \projectname{} allows \zynqps{} software to monitor stalls at a per-PC, per-cycle granularity.}
    \label{fig:08-improvement}
\end{figure}

After adding synthesizable stall counters to the \plshell{}, Figure~\ref{fig:08-improvement} shows the cycle-stack breakdown of stalls during execution of CoreMark~\cite{gal2012exploring}. While CoreMark is a flawed benchmark for full-system characterization, it is widely used as proof of microarchitectural optimization. Additionally, it is an ideal demonstration of performance optimization frameworks since there is so little low-hanging fruit remaining. Because \blackparrot{} is an in-order pipeline with large L1 caches, load-use stalls are a primary performance bottleneck, accounting for 18\% of stalls in CoreMark. Load-use stalls have two subtypes: load-arithmetic and load-control operations. For number crunching applications, load-arithmetic stalls prevent optimal operation of tight loops. For pointer chasing segments, load-control stalls add extra delays on every null check.

To reduce load-use stalls, we add a \textit{\catchup{}} ALU which is a secondary ALU located serially after the first ALU. Catch-up ALUs are a common way to improve performance in in-order cores. Out-of-order execution is often able to tolerate L1 hit latencies, so extra resources are better spent on more parallel ALUs for wider issue. For in-order cores, however, single threaded performance is sensitive to head-of-line blocking and so Catch-up ALUs can provide a substantial benefit. After justifying the idea in a high-level (Scale-Up) simulation model, we implement an RTL version of the idea in \projectname{} to evaluate marginal performance gains.

The \catchup{} ALU resides in EX2, parallel with the second stage of the D\$ access. When an integer or branch instruction has all dependencies met during issue, it is dispatched as normal to the Early ALU. Alternatively, when those dependencies are anticipated to be produced in EX2, the instruction is dispatched to the Catch-up ALU, which adds an additional cycle of latency, although fully-pipelined.

\begin{figure}
    \centering
    \resizebox{!}{0.5\columnwidth}{%
    \begin{tikzpicture}
    \node[right=2.5cm of n0] (n1) {EX};
    \node[right=2.5cm of n1] (n2) {MEM};
    \node[right=2.5cm of n2] (n3) {WB};

    \node[bsg_rectangle, minimum width=0.25cm, minimum height=2.5cm, below=0.50cm of n1.south] (n5) {};
    \node[bsg_rectangle, minimum width=0.25cm, minimum height=2.5cm, below=0.50cm of n2.south] (n6) {};
    \node[bsg_rectangle, minimum width=0.25cm, minimum height=2.5cm, below=0.50cm of n3.south] (n7) {};

    \node[bsg_trapezoid, minimum width=1.0cm, rotate=-90, above left=0.40cm and 0.75cm of n5] (n10) {};
    \draw ($(n10.south east) + (0.00cm, -0.15cm)$) -- ($(n10.south east) + (-0.50cm, -0.15cm)$);
    \node[bsg_trapezoid, dotted, minimum width=1.0cm, rotate=-90, above left=0.40cm and 0.75cm of n6] (n11) {};
    \node[bsg_trapezoid, minimum width=1.0cm, rotate=-90, above left=-0.10cm and 0.75cm of n7] (n12) {};

    \path [-]
        (n7.east) edge ($(n7.east) + (0.25cm, 0)$)
        ($(n7.east) + (0.25cm, 0)$) edge ($(n7.east) + (0.25cm, -1.50cm)$)
        ($(n7.east) + (0.25cm, -1.50cm)$) edge ($(n5.south west) - (1.00cm, 0.25cm)$)
        ($(n5.south west) - (1.00cm, 0.25cm)$) edge ($(n5.south west) - (1.00cm, -1.50cm)$)
        ($(n5.south west) - (1.00cm, -1.50cm)$) edge ($(n5.south west) - (0.80cm, -1.50cm)$)
        ;

    \path [-, dotted]
        (n7.east) edge ($(n7.east) + (0.25cm, 0)$)
        ($(n7.east) + (0.25cm, 0)$) edge ($(n7.east) + (0.25cm, -1.50cm)$)
        ($(n7.east) + (0.25cm, -1.50cm)$) edge ($(n6.south west) - (0.75cm, 0.25cm)$)
        ($(n6.south west) - (1.00cm, 0.25cm)$) edge ($(n6.south west) - (1.00cm, -1.50cm)$)
        ($(n6.south west) - (1.00cm, -1.50cm)$) edge ($(n6.south west) - (0.80cm, -1.50cm)$)
        ;

    \node[bsg_rectangle, minimum width=3.25cm, below right=0.50cm and 0.25cm of n5.east] (n13) {D\$};
    \node[bsg_trapezoid, minimum width=1.0cm, rotate=-90, above right=0.25cm and 0.25cm of n5.east] (n15) {};
    \path [-]
        ([xshift=0.25cm, yshift=-0.05cm]n5.east) edge[draw=white, thick] ([xshift=0.25cm, yshift=0.20cm]n5.east)
        ([xshift=0.25cm, yshift=0.20cm]n5.east) edge[draw=black] ([xshift=0.40cm, yshift=0.075cm]n5.east)
        ([xshift=0.40cm, yshift=0.075cm]n5.east) edge[draw=black] ([xshift=0.25cm, yshift=-0.05cm]n5.east)
        ;
    \node[bsg_trapezoid, dotted, minimum width=1.0cm, rotate=-90, above right=0.25cm and 0.25cm of n6.east] (n16) {};
    \path [-, dotted]
        ([xshift=0.25cm, yshift=-0.05cm]n6.east) edge[draw=white, solid, thick] ([xshift=0.25cm, yshift=0.20cm]n6.east)
        ([xshift=0.25cm, yshift=0.20cm]n6.east) edge[draw=black] ([xshift=0.40cm, yshift=0.075cm]n6.east)
        ([xshift=0.40cm, yshift=0.075cm]n6.east) edge[draw=black] ([xshift=0.25cm, yshift=-0.05cm]n6.east)
        ;

    \draw ([xshift=-0.15cm, yshift=-0.10cm]n15.west) -- ([xshift=-0.40cm, yshift=-0.10cm]n15.west);
    \draw ([xshift=-0.15cm, yshift=-0.60cm]n15.west) -- ([xshift=-0.40cm, yshift=-0.60cm]n15.west);
    \draw (n13.west) -- ([xshift=-0.25cm]n13.west);
    \draw[dotted] ([xshift=-0.15cm, yshift=-0.10cm]n16.west) -- ([xshift=-0.40cm, yshift=-0.10cm]n16.west);
    \draw[dotted] ([xshift=-0.15cm, yshift=-0.60cm]n16.west) -- ([xshift=-0.40cm, yshift=-0.60cm]n16.west);

    \draw ([xshift=0.15cm, yshift=0.35cm]n10.east) -- ([xshift=0.62cm, yshift=0.35cm]n10.east);
    \draw ([xshift=0.15cm, yshift=0.35cm]n11.east) -- ([xshift=0.62cm, yshift=0.35cm]n11.east);
    \draw ([xshift=0.15cm, yshift=0.35cm]n12.east) -- ([xshift=0.62cm, yshift=0.35cm]n12.east);

    \draw (n15) -- ($(n15) + (1.05cm, 0cm)$);
    \draw[dotted] (n16) -- ($(n16) + (1.05cm, 0cm)$);

    \path [-]
            (n13.east) edge ($(n13.east) + (0.25cm, 0cm)$)
            ($(n13.east) + (0.25cm, 0cm)$) edge ($(n13.east) + (0.25cm, 0.40cm)$)
            ($(n13.east) + (0.25cm, 0.40cm)$) edge ($(n13.east) + (0.45cm, 0.40cm)$)
        ;

    \path [-, dotted]
            ($(n12.east) + (0.15cm, 0.35cm)$) edge ($(n12.east) + (0.30cm, 0.35cm)$)
            ($(n12.east) + (0.30cm, 0.35cm)$) edge ($(n12.east) + (0.30cm, 1.95cm)$)
            ($(n12.east) + (0.30cm, 1.95cm)$) edge ($(n12.east) + (-2.85cm, 1.95cm)$)
            ($(n12.east) + (-2.85cm, 1.95cm)$) edge ($(n12.east) + (-2.85cm, 1.00cm)$)
            ($(n12.east) + (-2.85cm, 1.00cm)$) edge ($(n12.east) + (-2.65cm, 1.00cm)$)
        ;

    \path [-]
            ($(n12.east) + (0.15cm, 0.35cm)$) edge ($(n12.east) + (0.30cm, 0.35cm)$)
            ($(n12.east) + (0.30cm, 0.35cm)$) edge ($(n12.east) + (0.30cm, 1.95cm)$)
            ($(n12.east) + (0.30cm, 1.95cm)$) edge ($(n12.east) + (-5.35cm, 1.95cm)$)
            ($(n12.east) + (-5.35cm, 1.95cm)$) edge ($(n12.east) + (-5.35cm, 0.95cm)$)
            ($(n12.east) + (-5.35cm, 0.95cm)$) edge ($(n12.east) + (-5.15cm, 0.95cm)$)
        ;

    \path [-]
            ($(n11.east) + (0.15cm, 0.35cm)$) edge ($(n11.east) + (0.45cm, 0.35cm)$)
            ($(n11.east) + (0.45cm, 0.35cm)$) edge ($(n11.east) + (0.45cm, 1.30cm)$)
            ($(n11.east) + (0.45cm, 1.30cm)$) edge ($(n11.east) + (-2.75cm, 1.30cm)$)
            ($(n11.east) + (-2.75cm, 1.30cm)$) edge ($(n11.east) + (-2.75cm, 0.60cm)$)
            ($(n11.east) + (-2.75cm, 0.60cm)$) edge ($(n11.east) + (-2.65cm, 0.60cm)$)
        ;

    \draw ($(n15) + (1.05cm, 0cm)$) -- ($(n11.east) + (0.15cm, 0.35cm)$);

\end{tikzpicture}
    }
    \caption{A second \catchup{} ALU and set of bypass multiplexers allows the \catchup{} ALU to execute pipelined instructions. However, a dependent non-integer instruction following a \catchup{} operation will cause a bubble.}
    \label{fig:08-catchup}
\end{figure}
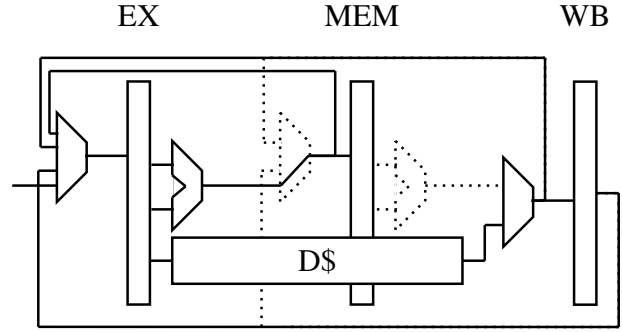

In addition to arithmetic operations, the Catch-up ALU also processes control flow instructions. Because RISC-V branch comparisons are easily transformed from existing subtraction and comparison operations, this support is cheap to add. However, this feature adds complexity to the handling of branch mispredictions. The \blackparrot{} pipeline resolves branches early in EX1 to reduce the misprediction penalty. In order for load-branch operations to take advantage of the \catchup{} ALU, the pipeline must suppress PC mismatches in EX1. Now, when the \catchup{} ALU detects a PC mismatch, the pipeline must be flushed in addition to redirecting the front-end. Therefore in \blackparrot{}, \catchup{} ALU mispredictions are treated as synchronous exceptions, reusing their mechanism for replaying and recovering state. Figure~\ref{fig:08-catchup} illustrates \catchup{} ALU modifications to a sample five-stage pipeline.

As shown in Figure~\ref{fig:08-catchup}, the \catchup{} ALU reduces load-use stalls from 43\% of stalls to 18\% of stalls,  resulting in an overall 4\% performance increase. There are additional stalls from dependencies on \catchup{} ALU instructions, which now have an additional cycle of latency. However, these extra stalls do not diminish the gains from optimizing the more common load-use case. Interestingly, branch-related stalls increase by 1.04\bsgtimes{}, as deeper speculation past EX1 triggers additional mispredictions. A further optimization could restrict speculation only to branches which are predicted strongly taken which would increase load-branch stalls but should reduce \catchup{} mispredictions. Leveraging cycle-accurate profiling with \projectname{} allows architects to easily identify potential bottlenecks as well as confirm both the positive and negative effects of their proposed improvements.

\subsection{\projectname{} ASIC Bring-up Boards}

\begin{figure}
    \centering
    \resizebox{0.60\columnwidth}{!}{%
    \begin{tikzpicture}
\node[bsg_rectangle, minimum width=6cm, minimum height=4.5cm] (n0) {};
\node[anchor=north, below=0.25cm of n0.north] {\textbf{FPGA Gateway Chip}};

\node[bsg_rectangle, minimum width=1.0cm, minimum height=1.0cm, below=0.90cm of n0.north] (n1) {\zynqps{}};
\node[right=1cm of n1.east] (n4) {Ethernet};
\node[left=1cm of n1.west] (n5) {DDR};
\node[bsg_rectangle, minimum width=5cm, below=0.50cm of n1.south] (n3) {\plshell{}};

\draw [->] (n1) edge (n4);
\draw [->] (n1) edge (n5);
\draw [<->] (n1) edge (n3);

\node[bsg_rectangle, minimum width=5cm, below=0.30cm of n3.south] (n6) {CSR and FIFO Bridges};

\draw [->] (n3) edge (n6);
\draw [<->] ([xshift=0.45cm]n6.south west) edge[ultra thick] ([xshift=0.45cm, yshift=-0.75cm]n6.south west);
\draw [<->] ([xshift=2.00cm]n6.south west) edge[ultra thick] ([xshift=2.00cm, yshift=-0.75cm]n6.south west);
\draw [<->] ([xshift=4.00cm]n6.south west) edge[ultra thick] ([xshift=4.00cm, yshift=-0.75cm]n6.south west);

\node[bsg_rectangle, minimum width=6cm, minimum height=1.25cm, below=0.25cm of n0.south] (n2) {\\\\\textbf{ASIC Test Chip}};
\draw (n2.south west) edge[draw=white] (n2.south east);
\draw (n2.north west) edge[xshift=0.25cm] ([yshift=0.25cm]n2.north west);
\draw (n2.north west) edge[xshift=0.50cm] ([yshift=0.25cm]n2.north west);
\draw (n2.north west) edge[xshift=0.75cm] ([yshift=0.25cm]n2.north west);
\draw (n2.north west) edge[xshift=1.00cm] ([yshift=0.25cm]n2.north west);
\draw (n2.north west) edge[xshift=1.25cm] ([yshift=0.25cm]n2.north west);
\draw (n2.north west) edge[xshift=1.50cm] ([yshift=0.25cm]n2.north west);
\draw (n2.north west) edge[xshift=1.75cm] ([yshift=0.25cm]n2.north west);
\draw (n2.north west) edge[xshift=2.00cm] ([yshift=0.25cm]n2.north west);
\draw (n2.north west) edge[xshift=2.25cm] ([yshift=0.25cm]n2.north west);
\draw (n2.north west) edge[xshift=2.50cm] ([yshift=0.25cm]n2.north west);
\draw (n2.north west) edge[xshift=2.75cm] ([yshift=0.25cm]n2.north west);
\draw (n2.north west) edge[xshift=3.00cm] ([yshift=0.25cm]n2.north west);
\draw (n2.north west) edge[xshift=3.25cm] ([yshift=0.25cm]n2.north west);
\draw (n2.north west) edge[xshift=3.50cm] ([yshift=0.25cm]n2.north west);
\draw (n2.north west) edge[xshift=3.75cm] ([yshift=0.25cm]n2.north west);
\draw (n2.north west) edge[xshift=4.00cm] ([yshift=0.25cm]n2.north west);
\draw (n2.north west) edge[xshift=4.25cm] ([yshift=0.25cm]n2.north west);
\draw (n2.north west) edge[xshift=4.50cm] ([yshift=0.25cm]n2.north west);
\draw (n2.north west) edge[xshift=4.75cm] ([yshift=0.25cm]n2.north west);
\draw (n2.north west) edge[xshift=5.00cm] ([yshift=0.25cm]n2.north west);
\draw (n2.north west) edge[xshift=5.25cm] ([yshift=0.25cm]n2.north west);
\draw (n2.north west) edge[xshift=5.50cm] ([yshift=0.25cm]n2.north west);
\draw (n2.north west) edge[xshift=5.75cm] ([yshift=0.25cm]n2.north west);

\node[bsg_rectangle, below right=0.10cm and 0.40cm of n2.north west] (n8) {JTAG};
\node[bsg_rectangle, right=0.15cm of n8.east] (n9) {DDR I/O};
\node[bsg_rectangle, right=0.15cm of n9.east] (n10) {SDDR DRAM};

\end{tikzpicture}
    }
    \caption{When taping out a chip verification is scoped sequentially, starting with C++ and RTL models and finishing with annotated gate-level simulations. Differences in simulation and bring-up environments prevent sharing infrastructure between pre- and post-silicon environments. \projectname{} unifies tape-in and tape-out  infrastructure, reducing maintenance times and accelerating bring-up.}
    \label{fig:08-case-usage-bringup}
\end{figure}
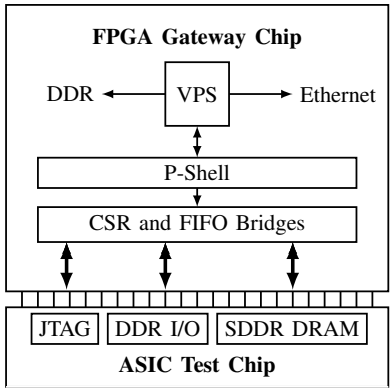

During its initial evolution, \projectname{} was also used as the design, prototype and bring-up infrastructure for a 14M gate, 28~nm ASIC developed by a boutique FPGA-based research and development firm. As a first generation test chip, operational mode fallbacks were essential. For instance, an experimental open-source LPDDR controller was taped out to accelerate applications. However, it was essential that the chip function enough to bring up all components independently. Additionally, it was not only possible but expected that during experiments on the chip to bring-up new subsystems, debug systemic issues and generate Shmoo plots~\cite{baker1997shmoo} for operational PPA, the chip will fail to respond, generate illegal traffic or otherwise operate out of specification.

As shown in Figure~\ref{fig:08-case-usage-bringup}, \projectname{} leverages the \plshell{} to allow bring-up software to interact with subsections of the DUT for large and slow gate-level simulations. Test chips are able to offload their memory controllers and I/O devices to the \zynqps{}, which allows pre-silicon bring-up to verify fallback functionality much earlier in the life-cycle. In the other direction, the flexible \plshell{} can connect to a wide variety of bitbanged configuration devices (JTAG, SPI, I2C), higher performance I/O links (CAN, UART, custom SERDES), and DRAM (LPDDR, SDDR, HBM). In addition to bringing up the actual chip with \projectname{}, users are able to substitute out fully detailed simulated execution models for faster-simulating smoke tests.

Using \projectname{}, the ASIC was able to pass initial smoke tests on the first day of bring-up despite power and packaging problems that prevented it from operating in normal voltage operations. By using non-blocking registers to bitbang and configure the attached test chip, the team was able to quickly iterate and explore modes without hanging the system, needing to re-flash the FPGA for different tests or debug the infrastructure itself.

\section{\panicroom{}: Ultra-Portable Baremetal Benchmarking}

Due to the complexity of benchmarking experimental processor designs, architects normalize performance across a wide range of applications such as ~\cite{henning2000spec,henning2006spec,bucek2018spec,bienia2008parsec}. There are several significant challenges associated with reusing the same applications for fabricated chips as for pre-silicon designs: notably, the scale of commercial benchmarks are incompatible with the massive slowdowns of RTL simulation. Prior works have proposed reducing input set size~\cite{kleinosowski2001adapting,todi2001speclite} and reducing instruction count through statistical sampling~\cite{hamerly2005simpoint,wenisch2006simflex}. Other approaches is creating targeted benchmarks which are intrinsically small and portable~\cite{gal2012exploring,guthaus2001mibench,pallister2013beebs}, but these have questionable correlation to high-performance microarchitectures.

\begin{figure}
    \centering
    \resizebox{0.9\columnwidth}{!}{%
    \newsavebox\codeboxmain
\newsavebox\codeboxshim

\begin{lrbox}{\codeboxmain}
\begin{lstlisting}[language=C++]
#include <stdio.h>
#include <stdlib.h>
#include <string.h>

int main() {
    // Read from a file
    FILE *hello = fopen("hello.txt");
    if (hello == NULL)
        return -1;

    while ((c = fgetc(hello)) != EOF)
    {
        putcharc(c);
    }

    fclose(hello);
    return 0;
}
\end{lstlisting}
\end{lrbox}

\begin{lrbox}{\codeboxshim}
\begin{lstlisting}
#include <stdlib.h>
#include <machine/panicroom_fs.h>

void panicroom_init(void);
void panicroom_exit(int exit_status);
void panicroom_sendchar(char ch);
int panicroom_getchar(void);
\end{lstlisting}
\end{lrbox}

\begin{tikzpicture}
\node[bsg_rectangle, fill=lightgray, font=\tiny, minimum width=1.50cm, minimum height=2.00cm                   ] (n0) {};
\node[font=\tiny, below=0.15cm   of n0.north] (t0) {Data};
\node[bsg_rectangle, fill=white, font=\tiny, minimum width=1.35cm, minimum height=1cm, above=0.00cm of n0] (n1) {LittleFS\\Disk Image};
\node[bsg_rectangle, fill=lightgray, font=\tiny, minimum width=1.50cm, minimum height=1.60cm, above=0cm of n0.north] (n2) {};
\node[font=\tiny, above=0.50cm of n2] (n3) {Newlib\\(C stdlib)};
\node[bsg_rectangle, fill=white, font=\tiny, minimum width=1.35cm, below=0.25cm of n2] (n3) {LittleFS\\Block Device};
\node[bsg_rectangle, fill=white, font=\tiny, minimum width=1.50cm, above=0.00cm of n2.north] (n4) {Platform Shim};
\node[bsg_rectangle, fill=lightgray, font=\tiny, minimum width=1.50cm, minimum height=1.95cm, above=0.00cm of n4.north] (n5) {User Program};

\node[bsg_rectangle, below right=0.00cm and 3.05cm of n5] (code0) {\resizebox{3.5cm}{!}{\usebox\codeboxmain}};
\node[bsg_rectangle, below=3.05cm of code0] (code1) {\resizebox{3.5cm}{!}{\usebox\codeboxshim}};

\draw [dotted] (n5.east) -- (code0.north west);
\draw [dotted] (n5.east) -- (code0.south west);

\draw [dotted] (n4.east) -- (code1.north west);
\draw [dotted] (n4.east) -- (code1.south west);

\draw [-] (n1.west) node[left=0.82cm of n1] (i0){} -- (i0.center);
\draw [-] (n3.west) node[left=0.82cm of n3] (i1){} -- (i1.center);
\draw [-] (i0.center) node[above=0.70cm of i0] (i2){} -- (i2.center);
\draw [-] (i1.center) -- (i2.center);
\draw [-] (i2.center) node[left=0.10cm of i2] (i3){} -- (i3.center);

\node [font=\tiny, rotate=90, left=0.10cm of i3] {I/O, DRAM};

\end{tikzpicture}
    }
    \caption{\panicroom{} provides filesystem and I/O operations to POSIX applications. Platform support needs 4 non-portable syscalls: init, exit, sendchar and getchar. All other syscall functionality is platform-independently provided by the \panicroom{} libgloss implementation. Programs cannot differentiate between \panicroom{} or a full OS, simply running benchmarks which otherwise require esoteric environments.}
    \label{fig:03-software-panicroom}
\end{figure}
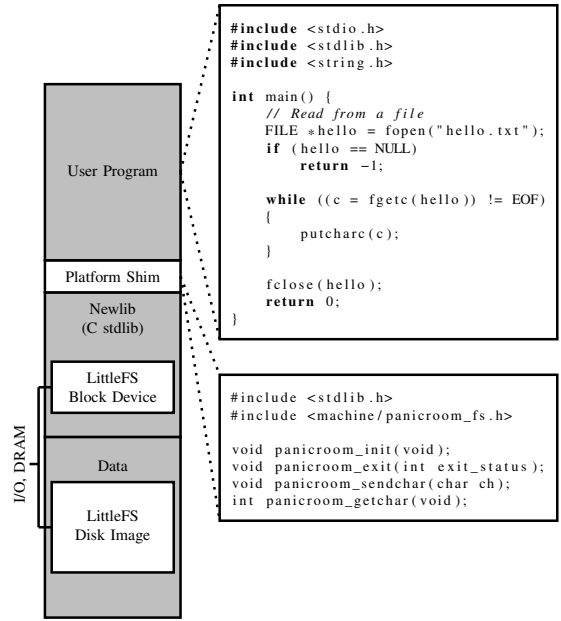

\begin{table}
    \centering
    \begin{threeparttable}
    \caption{\panicroom{} supports host functionality more portably than comparable proxy solutions. While a full debugging solution such as ARM Semi-Hosting provides an end-to-end experience, \panicroom{} is able to run identically in simulation and hardware, increasing verification correlation.}
    \label{tab-03-software:pcomparison}
    \small
\begin{tabular}{cccc}
\hline
\textbf{\begin{tabular}[c]{@{}c@{}}Proxy\\Solution\end{tabular}} & \textbf{\begin{tabular}[c]{@{}c@{}}Hardware\\Needed\tnote{0}\end{tabular}} & \textbf{\begin{tabular}[c]{@{}c@{}}LoC\\(Userspace)\tnote{1}\end{tabular}} & \textbf{\begin{tabular}[c]{@{}c@{}}Open-\\Source\end{tabular}} \\ \hline
RISCV-PK              & Host core  & 14157       & \checkmark       \\
RAW Interface         & Host core  & 6999        & \checkmark       \\
ARM Semi-Hosting      & Debugger   & -           &                  \\
\textbf{\panicroom{}} & DRAM       & \textbf{20} & \checkmark
\end{tabular}
    \begin{tablenotes}
        \footnotesize
        \item[0] While \projectname{} provides a \zynqps{} host, \panicroom{} can run fully untethered. 
        \item[1] We refer to non-benchmark, bare-metal code as "Userspace".
    \end{tablenotes}
    \end{threeparttable}
\end{table}

In addition to scale, commercial benchmarks also suffer from oversized scope. The most commonly evaluated suites rely on functions from the C standard library for I/O capability, filesystem operations and memory management. While it is interesting to evaluate the performance of an end-to-end system, during deep microarchitectural optimization architects often wish to observe bare-metal behavior. Yet, without operating system, it is impossible to run all but the most intentionally portable applications. Instead of glibc~\cite{glibc}, embedded systems typically rely on smaller stdlib implementations, but these lack necessary system call compatibility.

To bridge this gap, we introduce \textit{\panicroom{}}: a minimal, mostly-platform-agnostic C standard library implementation that enables running POSIX applications on bare-metal systems. \panicroom{} is built as a Board Support Package (BSP) on top of the light-weight C standard library newlib~\cite{newlib}. newlib elegantly separates system-specific functionality into an easily portable portion called libgloss. \panicroom{} implements the libgloss functionality using an open-source, lightweight, DRAM-based filesystem designed for embedded flash memories, ARM LittleFS~\cite{littlefs}.

As shown in Figure~\ref{fig:03-software-panicroom}, \panicroom{}{} implements file I/O system calls by translating them to LFS function calls, which in turn operate on memory. In contrast, proxy-based solutions such as RAW~\cite{waingold1997baring} and RISCV-PK~\cite{riscvpk} work by packaging I/O calls and tunneling to a host core, which executes the actual filesystem functionality. 

\panicroom{} eschews platform-dependent syscalls, whereas previous works require porting syscalls to open-source simulators, commercial simulators, FPGA emulation frameworks, ASIC test boards and PCIe hosted chips (see Table 2). A more subtle benefit is that transforming the I/O emulation from an asynchronous host interaction to a synchronous function, which makes execution deterministic and easily reproducible.

\section{High Fidelity Sampling with \projectname{}}
\begin{figure}
    \centering
    \includegraphics[width=\columnwidth]{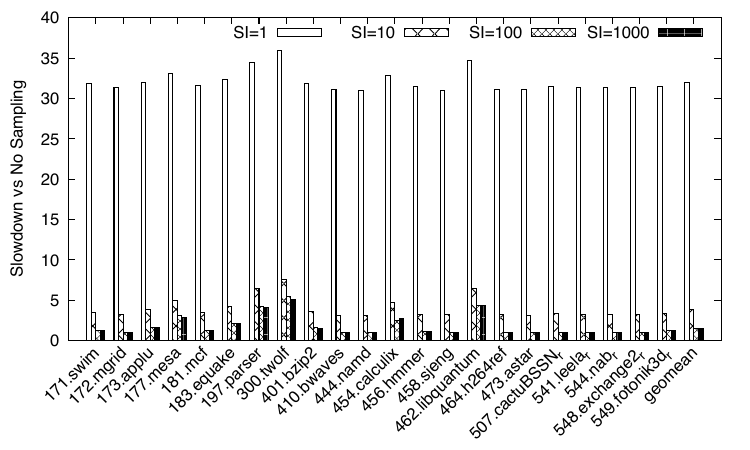}
    \caption{\projectname{} is able to dynamically switch co-emulation speed for sample rate. As sampling granularity decreases down to single step, there is a 32\bsgtimes{} slowdown. Therefore, best practice is to identify regions-of-interest and change sampling frequency to match importance.}
    \label{fig:10-slowdown}
\end{figure}

\label{sec:10-case-sampling}

While \projectname{} accelerates design emulation, validating and optimizing performance at a Scale-Down granularity additionally requires deep introspection. Unfortunately, FPGA-enabled acceleration of designs is famously opaque and extracting microarchitecural information is unintuitive. Traditional solutions include using vendor IP such as JTAG scan-chains or Xilinx ILA~\cite{xilinxila} to extract signals. However, these solutions are slow, proprietary, have a high area overhead, and operate out-of-band, therefore lacking capability to maintain cycle-accuracy at full bandwidth. In this section, we explore how to leverage \projectname{}'s sampling infrastructure to characterize \blackparrot{} through time-proportional~\cite{gottschall2023tea,gottschall2021tip} performance profiling.

To extract arbitrarily precise microarchitectural information from RTL, \projectname{} leverages the same clock-gating mechanism used for I/O co-emulation. Users instantiate a parameterizable number of synthesizable performance counters in the \plshell{}. These counters can be explicitly instantiated in the RTL, automatically generated by tools like FirePerf~\cite{fireperf}, or by hierarchically connecting PL counters to internal DUT signals. The latter does not require any modification to the DUT RTL and so is the simplest and least invasive solution. 

When profiling \blackparrot{}, the profiler annotates each cycle of execution with a PC and event classification (stall type or commit), attributing at the commit stage to maintain time-proportionality. The DUT streams samples to \zynqps{} across asynchronous FIFOs at a configurable sample rate, if necessary clock-gating identically to how \projectname{} manages emulation of interface timings. Critically, due to the backpressure mechanism, tuning profiling granularity becomes a simple trade-off between slowdown and precision. TEA~\cite{gottschall2023tea} and TIP~\cite{gottschall2021tip} have demonstrated the benefits of Oracular stall classification, but concluded that the bandwidth overhead is impractical. Figure~\ref{fig:10-slowdown} illustrates that with \projectname{}, an Oracle incurs only moderate overhead and enables unprecedented insight for performance debugging.

\begin{figure}
    \centering
    \includegraphics[width=\columnwidth]{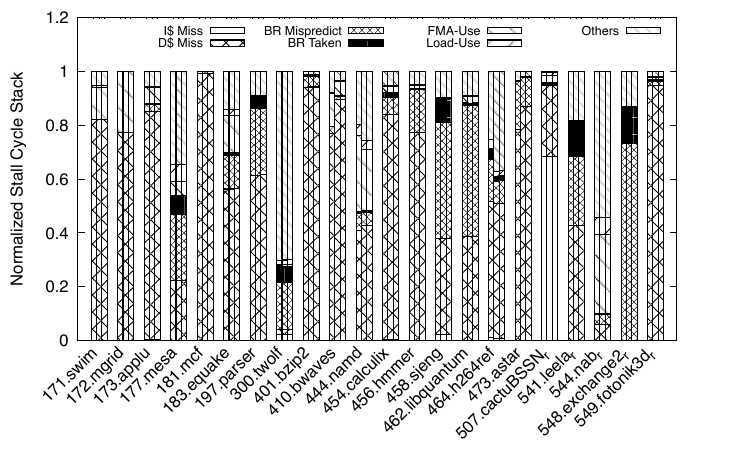}
    \caption{Stall stacks for sampling intervals 1, 10, and 100. Due to time-proportionality, stall stacks do not generally vary across sampling intervals. However, a few benchmarks such as 454.calculix and 464.h264ref have variances as high as 6.2\%. Oracular sampling through \projectname{} is able to accurately identify these stall sources.}
    \label{fig:10-breakdown}
\end{figure}

The \zynqps{} post-processes the stall information asynchronous to the DUT. Based on this information, profiler runtimes may chose to manipulate the sampling rate, perturb the DUT with emulated I/O traffic or monitor the runtime execution. Figure~\ref{fig:10-slowdown} shows the emulation slowdown for performance sampling of the core with error-free per-cycle sampling or with different sampling frequencies that results from DUT clock gating. Note that due to other clock gating factors (such as maintaining a memory timing model), with increasing sampling interval, the slowdown curve saturates to a different value based on the running benchmark. Due to the time-proportional nature of the profiler, stall stacks do not vary across sampling rates, as shown in Figure~\ref{fig:10-breakdown}. There are two resulting modalities for performance profiling in \projectname{}: running a coarse-grained regression suite to gain a sense of important stall categories, and running a fine-grained analysis to produce Oracular stall attributions to individual PCs.

\section{Scalably Acclerating Coverage Collection with \projectname{}}

Code coverage is paramount in pre-silicon verification~\cite{brahme1984functional}. Because of the enormous complexity of a large design, any simulation-based approach is merely sampling its state space. Still, state-of-art functional verification flows rely on coverpoints and covergroups to identify testing gaps. Recently, slow RTL simulation speeds have motivated verification teams to accelerate coverage collection using FPGAs. \projectname{} supports automated, low-overhead coverage extraction through synthesizable coverpoints. In this section, we explore the design space and methodology for monitoring and extracting Mux Toggle coverage introduced by RFUZZ~\cite{rfuzz} using \projectname{}\footnote{We exclude full cross-covergroup tracking which would assess \textit{all} control paths in the hardware, as the state space quickly becomes unmanageable.}.

\begin{figure}
    \centering
    \includegraphics[width=\columnwidth]{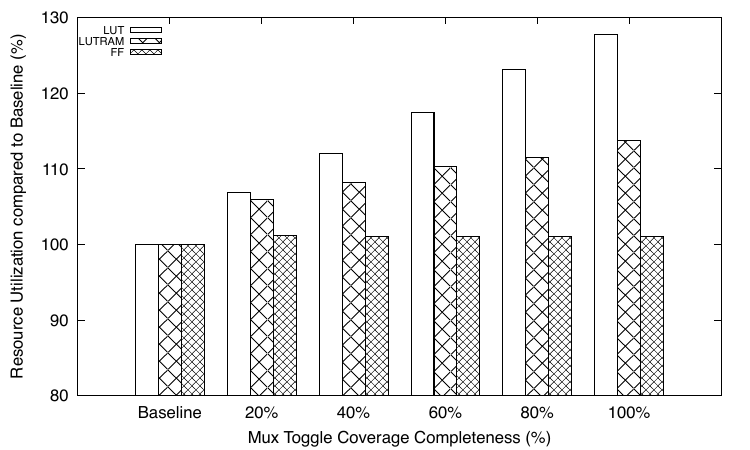}
    \caption{All 3284 Mux Toggle coverpoints for \blackparrot{} consume an extra 13\% LUTs and 27\% FFs. If the overhead is unacceptable, designers may use \farm{} clusters to parallelize coverpoints and lower per-board overheads.}
    \label{fig:11-case-coverage}
\end{figure}


\projectname{} coverpoints comprise all mux select signals, based on the expectation that the cumulative select-signal toggles correlate with the exercising of all individual control paths in the hardware. To automate identification of the coverpoints from the design RTL, we employ a SystemVerilog parser, Surelog~\cite{surelog}, to first generate an abstract syntax tree representation of the elaborated design, and then execute a customizable visitor at the root of the tree. Our visitor accumulates SystemVerilog structural elements that are likely to translate into Mux select-signals: the conditions in branch statements and ternary operations, while intelligently ignoring static expressions. The visitor also identifies case statement conditions for extracting FSM coverage for a more holistic coverage assessment. 

Once identified, coverpoints are implemented as single-bit registers, unlike previous hardware coverpoint implementations~\cite{sic} which use saturating counters to track toggles. Figure~\ref{fig:11-case-coverage} shows the incremental overheads of instrumenting the \blackparrot{} processor, a relatively control-heavy design. As a result, \projectname{} is able to obtain finer-grained coverage increments efficiently at the cost of under-representing coverage, thus enabling coverage-guided fuzzing. In most automated verification environments, under-representation is preferable to over-representation, as additional runtime can fill in the omitted toggles. Moreover, incremental coverage can be profiled dynamically when tracking coverage throughout program execution. Future fuzzers may take advantage of this interactivity in \projectname{} through early termination and directed mutation of test programs.

\section{Related Work}

\begin{table*}
    \centering
    \begin{threeparttable}
    \caption{Compared to other FPGA emulation Platforms, \projectname{} has minimal TCO for the smallest designs. Aside from enabling the use of low-end FPGAs, the lack of a dedicated host server reduces hardware cost and eliminates sysadmin overheads. Additionally, by eliminating dependencies on vendor IP, \projectname{} is able to provide cycle-accurate co-simulation using open-source tools at no licensing cost to the user.}
    \label{tab-96-related:comparison}
    \small
\begin{tabular}{ccccccccc}
\hline
\textbf{Platform} &
  \textbf{Cost Model} &
  \textbf{Design Ratio~\tnote{0}} &
  \textbf{Host} &
  \textbf{\begin{tabular}[c]{@{}c@{}}Cycle-\\ Accurate\end{tabular}} &
  \textbf{\begin{tabular}[c]{@{}c@{}}Co-\\ Simulation~\tnote{1}\end{tabular}} &
  \textbf{Coverage} &
  \textbf{\begin{tabular}[c]{@{}c@{}}Vendor-\\ Agnostic\end{tabular}} &
  \textbf{\begin{tabular}[c]{@{}c@{}}Open-\\ Source\end{tabular}} \\ \hline
Commercial~\tnote{2} & High-End Cluster & 1:1 & Proprietary & \checkmark{} & \checkmark{} & \checkmark{} &              &              \\
FireSim~\cite{firesim}              & Cloud Rental     & N:N & AWS F1      & \checkmark{} &              & \checkmark{} &              & \checkmark{} \\
SMAPPIC~\cite{chirkov2023smappic}              & Cloud Rental     & N:1 & AWS F1      &              &              &              &              & \checkmark{} \\
FreezeTime~\cite{mosanu2023freezetime}           & High-End Board   & 1=1 & PCIe Server & \checkmark{} &              &              & \checkmark{} & \\
\textbf{Condominium} &
  \textbf{\begin{tabular}[c]{@{}c@{}}Low-End Board/\\ Low-End Cluster\end{tabular}} &
  \textbf{N=N} &
  \textbf{None} &
  \textbf{\checkmark{}} &
  \textbf{\checkmark{}} &
  \textbf{\checkmark{}} &
  \textbf{\checkmark{}} &
  \textbf{\checkmark{}}
\end{tabular}
    \begin{tablenotes}
        \footnotesize
        \item[0] The ratio of FPGAs:Designs in a single system emulation. Commercial tools map large hierarchies into large clusters. Firesim is able to emulate arbitrarily large systems using cloud auto-scaling. SMAPPIC is able to split large designs across FPGAs. FreezeTime maps a single design to a single FPGA. Finally, Condominium maps a number of designs to a fixed-sized local cluster.
        \item[1] Co-simulation refers to the ability to reproduce the cycle-exact output of a system emulation on an RTL simulator such as Verilator~\cite{verilator} (albeit at significant slowdown). This ability is essential in emulation-system debugging.
        \item[2] We combine Synopsys Zebu~\cite{zebu}, Cadence Palladium~\cite{palladium} and Mentor Veloce~\cite{veloce} with similar features and limitations.
    \end{tablenotes}
    \end{threeparttable}
\end{table*}

While \zynqparrot shares similarities with many FPGA-accelerated prototyping platforms, its Scale-Down focus and aggressive portability make it uniquely cost and effort effective. In this section, we compare to existing projects which offer subsets of the features in \zynqparrot.

\subsection{Gate-Level Accelerated Emulation}
Teams desiring a turnkey solution to RTL emulation employ commercial tools for FPGA-accelerated design modeling, such as Cadence Palladium~\cite{palladium}, Synopsys Zebu~\cite{zebu} and Mentor Veloce~\cite{veloce}. Unfortunately this convenience is costly, with obfuscated pricing up to millions of dollars. In contrast \zynqparrot is free and open-source, with an initial required investment up to hundreds of dollars.

\subsection{Emulating Large Systems with FPGAs}
FireSim~\cite{firesim}, DIABLO~\cite{tan2015diablo} and SMAPPIC~\cite{chirkov2023smappic} focus on scaling up emulations to analyze large-scale designs such as datacenter-scale systems. They work by partitioning the system design over multiple FPGAs and using Ethernet-based token-passing systems to capture inter-node timing. Because they are based on AWS F1~\cite{aws} infrastructure, the emulation model relies on proprietary vendor libraries for the hardened AWS shell as well as PCIe DMA interfaces. 
As \zynqparrot focuses on single-node systems, it allows for local execution with open-source simulations, resulting in a much lower recurring cost. In contrast, a local version for a comparable F1 FPGA setup, may cost tens of thousands of dollars.

\subsection{Decomposed FPGA emulation}
Similar to a Scale-Down methodology, Protoflex~\cite{chung2009protoflex} and FAST~\cite{chiou2007fpga} accelerate performance analysis using FPGAs. However, they focus on acceleration of large, slow, cycle-accurate models, attempting to gain performance insights into systems too large to simulate in a reasonable time frame. In contrast, \zynqparrot allows for cycle-accurate emulation of arbitrary RTL so that architects can easily validate and debug performance with the deep introspection that RTL provides.

FreezeTime~\cite{mosanu2023freezetime} uses time multiplexing for architectural virtualization of system components. Similar to \zynqparrot, Freezetime leverages BUFGCE FPGA primitives to stall emulated blocks while virtualized blocks process cycle-accurate timing models. However, \zynqparrot achieves greater flexibility and lower resource overheads by executing standard C++ timing models in the \plshell rather than custom control logic per virtualized interface.

\subsection{FPGA-Accelerated Performance Analysis}
While custom cycle-level simulators and silicon performance counters are state-of-art for commercial performance validation, researchers have also proposed using accelerated sampling for microarchitectural debugging.

FirePerf~\cite{fireperf} provides two categories of microarchitectural analysis: commit tracing via TraceRV and out-of-band hardware profiling via AutoCounters. \zynqparrot supports not only commit tracing and out-of-band event counters via \plshell CSRs but also dispatch-time stall tracing, allowing for deeper debugging insights. Additionally, \zynqparrot is written in standard SystemVerilog rather than Chisel~\cite{chisel}, making it more familiar to hardware designers. 

TEA~\cite{gottschall2023tea} and TIP~\cite{gottschall2021tip} propose time-proportional event analysis by creating Per-Instruction Cycle Stacks (\textit{PICS}) to unify performance profiling and performance event analysis. While TEA and TIP are able to accurately ascribe microarchitectural events on average, they rely on statistical sampling by periodically interrupting the program that disrupts non-interference. Because \zynqparrot combines commit-stage cycle attribution with cycle accurate tracing, it is able to accurately attribute stalls without any sampling error, as well as trade co-emulation speed for sampling accuracy.

\subsection{FPGA-based Coverage Collection}

FirePerf~\cite{fireperf} injects synthesizable coverpoints and then extracts coverage through a scan-chain. Instead, \zynqparrot automatically instruments  designs and extracts stall data through the \plshell, without dedicated scan hardware. Simulator Independent Coverage~\cite{sic} introduces a flow for injecting hardware coverpoints into Chisel~\cite{chisel} designs through the introduction of a new \textit{cover} keyword. In contrast, \zynqparrot coverage instrumentation uses standard SystemVerilog primitives.

\section{Conclusion}

We present \projectname{}, a Scale-Down FPGA-based modelling platform capable of non-interfering, cycle-accurate co-emulations of arbitrary RTL designs. Vendor-agnostic and fully open-source, \projectname{} provides architects with a accurate, convenient and low-cost infrastructure to prototype designs. \projectname{} is a cheaper alternative to FPGA cloud infrastructures for small-scale experiments and provides vendor-agnosticism. An iterative Scale-Down/Scale-Up methodology allows architects to focus on subtle microarchitectural optimizations and avoid re-analysis of issues that only exist at either scale.

In this work we have explored diverse use-cases for \projectname{}: distributed acceleration for a class of architecture students, performance debugging, functional verification and tapeout bringup for a complex 28~nm SoC. These studies are meant to illustrate \projectname{}'s fitness for teaching, research and industrial development. We examined instrastructure enablements: ultra fine-grained sampling and hybrid FPGA-software coverpoint strategies. We believe the \projectname{}, the \plshell{} interface and its open-source co-simulation libraries will accelerate RTL prototyping across teams and fields, helping develop better architectures faster.


\bibliographystyle{IEEEtranS}
\bibliography{refs}

\begin{thebibliography}{10}
\providecommand{\url}[1]{#1}
\csname url@samestyle\endcsname
\providecommand{\newblock}{\relax}
\providecommand{\bibinfo}[2]{#2}
\providecommand{\BIBentrySTDinterwordspacing}{\spaceskip=0pt\relax}
\providecommand{\BIBentryALTinterwordstretchfactor}{4}
\providecommand{\BIBentryALTinterwordspacing}{\spaceskip=\fontdimen2\font plus
\BIBentryALTinterwordstretchfactor\fontdimen3\font minus
  \fontdimen4\font\relax}
\providecommand{\BIBforeignlanguage}[2]{{%
\expandafter\ifx\csname l@#1\endcsname\relax
\typeout{** WARNING: IEEEtranS.bst: No hyphenation pattern has been}%
\typeout{** loaded for the language `#1'. Using the pattern for}%
\typeout{** the default language instead.}%
\else
\language=\csname l@#1\endcsname
\fi
#2}}
\providecommand{\BIBdecl}{\relax}
\BIBdecl

\bibitem{alibaba}
Alibaba, ``https://www.alibabacloud.com/product/computing,'' 2023.

\bibitem{surelog}
C.~Alliance, ``https://github.com/chipsalliance/surelog,'' 2023.

\bibitem{aws}
Amazon, ``Amazon web services. 2022. amazon ec2 f1 instances.
  https://aws.amazon.com/ec2/instance-types/f1/,'' 2023.

\bibitem{awspricing}
Amazon, ``https://aws.amazon.com/ec2/spot/pricing/,'' 2023.

\bibitem{renode}
AntMicro, ``https://github.com/renode/renode,'' 2023.

\bibitem{armaxi}
ARM,
  ``https://developer.arm.com/documentation/102202/0300/axi-protocol-overview,''
  2023.

\bibitem{littlefs}
ARM, ``https://github.com/littlefs-project/littlefs,'' 2023.

\bibitem{xilinxila}
K.~Arshak, E.~Jafer, and C.~Ibala, ``Testing fpga based digital system using
  xilinx chipscope logic analyzer,'' in \emph{2006 29th International Spring
  Seminar on Electronics Technology}.\hskip 1em plus 0.5em minus 0.4em\relax
  IEEE, 2006, pp. 355--360.

\bibitem{ultra96pricing}
AVnet,
  ``https://www.avnet.com/wps/portal/us/products/avnet-boards/avnet-board-families/ultra96-v2/,''
  2023.

\bibitem{chisel}
\BIBentryALTinterwordspacing
J.~Bachrach, H.~Vo, B.~Richards, Y.~Lee, A.~Waterman, R.~Avi\v{z}ienis,
  J.~Wawrzynek, and K.~Asanovi\'{c}, ``Chisel: Constructing hardware in a scala
  embedded language,'' in \emph{Proceedings of the 49th Annual Design
  Automation Conference}, ser. DAC '12.\hskip 1em plus 0.5em minus 0.4em\relax
  New York, NY, USA: Association for Computing Machinery, 2012, p. 1216–1225.
  [Online]. Available: \url{https://doi.org/10.1145/2228360.2228584}
\BIBentrySTDinterwordspacing

\bibitem{baker1997shmoo}
K.~Baker and J.~Van~Beers, ``Shmoo plotting: The black art of ic testing,''
  \emph{IEEE Design \& Test of Computers}, vol.~14, no.~3, pp. 90--97, 1997.

\bibitem{bienia2008parsec}
C.~Bienia, S.~Kumar, J.~P. Singh, and K.~Li, ``The parsec benchmark suite:
  Characterization and architectural implications,'' in \emph{Proceedings of
  the 17th international conference on Parallel architectures and compilation
  techniques}, 2008, pp. 72--81.

\bibitem{4785534}
M.~Bohr, ``A 30 year retrospective on dennard's mosfet scaling paper,''
  \emph{IEEE Solid-State Circuits Society Newsletter}, vol.~12, no.~1, pp.
  11--13, 2007.

\bibitem{brahme1984functional}
Brahme and Abraham, ``Functional testing of microprocessors,'' \emph{IEEE
  transactions on Computers}, vol. 100, no.~6, pp. 475--485, 1984.

\bibitem{bucek2018spec}
J.~Bucek, K.-D. Lange, and J.~v.~Kistowski, ``Spec cpu2017: Next-generation
  compute benchmark,'' in \emph{Companion of the 2018 ACM/SPEC International
  Conference on Performance Engineering}, 2018, pp. 41--42.

\bibitem{palladium}
Cadence,
  ``https://www.cadence.com/en\_us/home/tools/system-design-and-verification/emulation-and-prototyping/palladium.html,''
  2023.

\bibitem{chiou2007fpga}
D.~Chiou, D.~Sunwoo, J.~Kim, N.~A. Patil, W.~Reinhart, D.~E. Johnson, J.~Keefe,
  and H.~Angepat, ``Fpga-accelerated simulation technologies (fast): Fast,
  full-system, cycle-accurate simulators,'' in \emph{40th Annual IEEE/ACM
  International Symposium on Microarchitecture (MICRO 2007)}.\hskip 1em plus
  0.5em minus 0.4em\relax IEEE, 2007, pp. 249--261.

\bibitem{dromajo}
ChipsAlliance, ``https://github.com/chipsalliance/dromajo,'' 2023.

\bibitem{chirkov2023smappic}
G.~Chirkov and D.~Wentzlaff, ``Smappic: Scalable multi-fpga architecture
  prototype platform in the cloud,'' in \emph{Proceedings of the 28th ACM
  International Conference on Architectural Support for Programming Languages
  and Operating Systems, Volume 2}, 2023, pp. 733--746.

\bibitem{chung2009protoflex}
E.~S. Chung, M.~K. Papamichael, E.~Nurvitadhi, J.~C. Hoe, K.~Mai, and
  B.~Falsafi, ``Protoflex: Towards scalable, full-system multiprocessor
  simulations using fpgas,'' \emph{ACM Transactions on Reconfigurable
  Technology and Systems (TRETS)}, vol.~2, no.~2, pp. 1--32, 2009.

\bibitem{raspberrypi}
R.~P. Foundation, ``https://www.raspberrypi.org,'' 2023.

\bibitem{riscvpk}
R.-V. Foundation, ``https://github.com/riscv-software-src/riscv-pk,'' 2023.

\bibitem{newlib}
FSF, ``https://sourceware.org/newlib/,'' 2023.

\bibitem{glibc}
FSF, ``https://www.gnu.org/software/libc/,'' 2023.

\bibitem{gal2012exploring}
S.~Gal-On and M.~Levy, ``Exploring coremark a benchmark maximizing simplicity
  and efficacy,'' \emph{The Embedded Microprocessor Benchmark Consortium},
  2012.

\bibitem{gottschall2021tip}
B.~Gottschall, L.~Eeckhout, and M.~Jahre, ``Tip: Time-proportional instruction
  profiling,'' in \emph{MICRO-54: 54th Annual IEEE/ACM International Symposium
  on Microarchitecture}, 2021, pp. 15--27.

\bibitem{gottschall2023tea}
B.~Gottschall, L.~Eeckhout, and M.~Jahre, ``Tea: Time-proportional event
  analysis,'' in \emph{Proceedings of the 50th Annual International Symposium
  on Computer Architecture}, 2023, pp. 1--13.

\bibitem{guthaus2001mibench}
M.~R. Guthaus, J.~S. Ringenberg, D.~Ernst, T.~M. Austin, T.~Mudge, and R.~B.
  Brown, ``Mibench: A free, commercially representative embedded benchmark
  suite,'' in \emph{Proceedings of the fourth annual IEEE international
  workshop on workload characterization. WWC-4 (Cat. No. 01EX538)}.\hskip 1em
  plus 0.5em minus 0.4em\relax IEEE, 2001, pp. 3--14.

\bibitem{hamerly2005simpoint}
G.~Hamerly, E.~Perelman, J.~Lau, and B.~Calder, ``Simpoint 3.0: Faster and more
  flexible program phase analysis,'' \emph{Journal of Instruction Level
  Parallelism}, vol.~7, no.~4, pp. 1--28, 2005.

\bibitem{henning2000spec}
J.~L. Henning, ``Spec cpu2000: Measuring cpu performance in the new
  millennium,'' \emph{Computer}, vol.~33, no.~7, pp. 28--35, 2000.

\bibitem{henning2006spec}
J.~L. Henning, ``Spec cpu2006 benchmark descriptions,'' \emph{ACM SIGARCH
  Computer Architecture News}, vol.~34, no.~4, pp. 1--17, 2006.

\bibitem{karandikar2018firesim}
S.~Karandikar, H.~Mao, D.~Kim, D.~Biancolin, A.~Amid, D.~Lee, N.~Pemberton,
  E.~Amaro, C.~Schmidt, A.~Chopra \emph{et~al.}, ``Firesim: Fpga-accelerated
  cycle-exact scale-out system simulation in the public cloud,'' in \emph{2018
  ACM/IEEE 45th Annual International Symposium on Computer Architecture
  (ISCA)}.\hskip 1em plus 0.5em minus 0.4em\relax IEEE, 2018, pp. 29--42.

\bibitem{firesim}
\BIBentryALTinterwordspacing
S.~Karandikar, H.~Mao, D.~Kim, D.~Biancolin, A.~Amid, D.~Lee, N.~Pemberton,
  E.~Amaro, C.~Schmidt, A.~Chopra, Q.~Huang, K.~Kovacs, B.~Nikolic, R.~Katz,
  J.~Bachrach, and K.~Asanovi\'{c}, ``{FireSim}: {FPGA}-accelerated cycle-exact
  scale-out system simulation in the public cloud,'' in \emph{Proceedings of
  the 45th Annual International Symposium on Computer Architecture}, ser. ISCA
  '18.\hskip 1em plus 0.5em minus 0.4em\relax Piscataway, NJ, USA: IEEE Press,
  2018, pp. 29--42. [Online]. Available:
  \url{https://doi.org/10.1109/ISCA.2018.00014}
\BIBentrySTDinterwordspacing

\bibitem{fireperf}
\BIBentryALTinterwordspacing
S.~Karandikar, A.~Ou, A.~Amid, H.~Mao, R.~Katz, B.~Nikoli\'{c}, and
  K.~Asanovi\'{c}, ``Fireperf: Fpga-accelerated full-system hardware/software
  performance profiling and co-design,'' in \emph{Proceedings of the
  Twenty-Fifth International Conference on Architectural Support for
  Programming Languages and Operating Systems}, ser. ASPLOS '20.\hskip 1em plus
  0.5em minus 0.4em\relax New York, NY, USA: Association for Computing
  Machinery, 2020, p. 715–731. [Online]. Available:
  \url{https://doi.org/10.1145/3373376.3378455}
\BIBentrySTDinterwordspacing

\bibitem{kleinosowski2001adapting}
A.~KleinOsowski, J.~Flynn, N.~Meares, and D.~J. Lilja, ``Adapting the spec 2000
  benchmark suite for simulation-based computer architecture research,''
  \emph{Workload characterization of emerging computer applications}, pp.
  83--100, 2001.

\bibitem{sic}
\BIBentryALTinterwordspacing
K.~Laeufer, V.~Iyer, D.~Biancolin, J.~Bachrach, B.~Nikoli\'{c}, and K.~Sen,
  ``Simulator independent coverage for rtl hardware languages,'' in
  \emph{Proceedings of the 28th ACM International Conference on Architectural
  Support for Programming Languages and Operating Systems, Volume 3}, ser.
  ASPLOS 2023.\hskip 1em plus 0.5em minus 0.4em\relax New York, NY, USA:
  Association for Computing Machinery, 2023, p. 606–615. [Online]. Available:
  \url{https://doi.org/10.1145/3582016.3582019}
\BIBentrySTDinterwordspacing

\bibitem{rfuzz}
K.~Laeufer, J.~Koenig, D.~Kim, J.~Bachrach, and K.~Sen, ``Rfuzz:
  Coverage-directed fuzz testing of rtl on fpgas,'' in \emph{2018 IEEE/ACM
  International Conference on Computer-Aided Design (ICCAD)}, 2018, pp. 1--8.

\bibitem{veloce}
Mentor, ``https://eda.sw.siemens.com/en-us/ic/veloce/,'' 2023.

\bibitem{mirasol2019scaledown}
F.~Mirasol, ``The principle of scaling down,'' \emph{BioPharm International
  32}, 2019.

\bibitem{mosanu2023freezetime}
S.~Mosanu, J.~Fixelle, M.~N. Sakib, K.~Skadron, and M.~Stan, ``Freezetime:
  Towards system emulation through architectural virtualization,'' in
  \emph{2023 IEEE International Parallel and Distributed Processing Symposium
  Workshops (IPDPSW)}.\hskip 1em plus 0.5em minus 0.4em\relax IEEE, 2023, pp.
  129--136.

\bibitem{pallister2013beebs}
J.~Pallister, S.~Hollis, and J.~Bennett, ``Beebs: Open benchmarks for energy
  measurements on embedded platforms,'' \emph{arXiv preprint arXiv:1308.5174},
  2013.

\bibitem{petrisko2020blackparrot}
D.~Petrisko, F.~Gilani, M.~Wyse, D.~C. Jung, S.~Davidson, P.~Gao, C.~Zhao,
  Z.~Azad, S.~Canakci, B.~Veluri \emph{et~al.}, ``Blackparrot: An agile
  open-source risc-v multicore for accelerator socs,'' \emph{IEEE Micro},
  vol.~40, no.~4, pp. 93--102, 2020.

\bibitem{verilator}
W.~Snyder, ``https://github.com/verilator/verilator,'' 2024.

\bibitem{zebu}
Synopsys, ``https://www.synopsys.com/verification/emulation/zebu-server.html,''
  2023.

\bibitem{tan2015diablo}
Z.~Tan, Z.~Qian, X.~Chen, K.~Asanovic, and D.~Patterson, ``Diablo: A
  warehouse-scale computer network simulator using fpgas,'' \emph{ACM SIGPLAN
  Notices}, vol.~50, no.~4, pp. 207--221, 2015.

\bibitem{taylor2018basejump}
M.~B. Taylor, ``Basejump stl: Systemverilog needs a standard template library
  for hardware design,'' in \emph{Proceedings of the 55th Annual Design
  Automation Conference}, 2018, pp. 1--6.

\bibitem{todi2001speclite}
R.~Todi, ``Speclite: using representative samples to reduce spec cpu2000
  workload,'' in \emph{Proceedings of the Fourth Annual IEEE International
  Workshop on Workload Characterization. WWC-4 (Cat. No. 01EX538)}.\hskip 1em
  plus 0.5em minus 0.4em\relax IEEE, 2001, pp. 15--23.

\bibitem{pynqz2}
TUL, ``https://www.tulembedded.com/fpga/productspynq-z2.html,'' 2023.

\bibitem{waingold1997baring}
E.~Waingold, M.~Taylor, D.~Srikrishna, V.~Sarkar, W.~Lee, V.~Lee, J.~Kim,
  M.~Frank, P.~Finch, R.~Barua \emph{et~al.}, ``Baring it all to software: Raw
  machines,'' \emph{Computer}, vol.~30, no.~9, pp. 86--93, 1997.

\bibitem{wenisch2006simflex}
T.~F. Wenisch, R.~E. Wunderlich, M.~Ferdman, A.~Ailamaki, B.~Falsafi, and J.~C.
  Hoe, ``Simflex: statistical sampling of computer system simulation,''
  \emph{IEEE Micro}, vol.~26, no.~4, pp. 18--31, 2006.

\bibitem{wong2017superscalar}
H.~T.-H. Wong, \emph{A superscalar out-of-order x86 soft processor for
  fpga}.\hskip 1em plus 0.5em minus 0.4em\relax University of Toronto (Canada),
  2017.

\bibitem{ug116}
Xilinx, ``Device reliability report (ug116),'' Xilinx, Tech. Rep., 2023.

\bibitem{xdma}
Xilinx, ``https://docs.xilinx.com/r/en-us/pg195-pcie-dma,'' 2023.

\bibitem{petalinux}
Xilinx,
  ``https://docs.xilinx.com/r/en-us/ug1144-petalinux-tools-reference-guide/introduction,''
  2023.

\bibitem{zynq}
Xilinx,
  ``https://www.xilinx.com/products/boards-and-kits/device-family/nav-zynq-7000.html,''
  2023.

\bibitem{pynq}
Xilinx, ``http://www.pynq.io/board.html,'' 2023.

\end{thebibliography}

\end{document}